\documentclass[a4paper,journal]{IEEEtran}
\usepackage{supertabular} 
\usepackage{graphicx}
\usepackage{epstopdf}
\usepackage{placeins}
\usepackage{array}
\DeclareGraphicsExtensions{.eps,.pdf} 
\graphicspath{ {figures/} }
\usepackage{cite}
\usepackage{amssymb,amsmath}
\usepackage{amssymb,amsmath,mathtools}
\usepackage{amssymb,amsmath}
\usepackage{algorithm,algpseudocode}
\usepackage{color}
\usepackage{multirow}
\usepackage{stfloats}
\usepackage{footmisc}
\usepackage[table, svgnames, dvipsnames]{xcolor}
\usepackage{makecell, cellspace, caption}

\newcolumntype{P}[1]{>{\centering\arraybackslash}p{#1}}

\usepackage{subfig}
\makeatletter
\newcommand\restartchapters{\par
  \setcounter{chapter}{0}%
  \setcounter{section}{0}%
  \gdef\@chapapp{\chaptername}%
  \gdef\thechapter{\@arabic\c@chapter}}
\makeatother

\begin{document}

\algnewcommand\algorithmicswitch{\textbf{switch}}
\algnewcommand\algorithmiccase{\textbf{case}}
\algnewcommand\algorithmicassert{\texttt{assert}}
\algnewcommand\Assert[1]{\State \algorithmicassert(#1)}
\algrenewcommand{\alglinenumber}[1]{\scriptsize#1:}

\algdef{SE}[SWITCH]{Switch}{EndSwitch}[1]{\algorithmicswitch\ #1\ \algorithmicdo}{\algorithmicend\ \algorithmicswitch}%
\algdef{SE}[CASE]{Case}{EndCase}[1]{\algorithmiccase\ #1}{\algorithmicend\ \algorithmiccase}%
\algtext*{EndCase}%

\bstctlcite{IEEEexample:BSTcontrol}
\title{NB-IoT Random Access for Non-Terrestrial Networks:  {Preamble Detection and Uplink Synchronization}}
\author{
\IEEEauthorblockN{Houcine Chougrani,  Steven Kisseleff,~\IEEEmembership{Member,~IEEE}, Wallace A. Martins,~\IEEEmembership{Senior Member,~IEEE},\\ and Symeon Chatzinotas},~\IEEEmembership{Senior Member,~IEEE}
 \thanks{H. Chougrani,  S. Kisseleff, W.A. Martins, and S. Chatzinotas are with the Interdisciplinary Centre for Security, Reliability and Trust (SnT), University of Luxembourg, Luxembourg. W.A. Martins is also with the Federal University of Rio de Janeiro (UFRJ), Brazil. E-mails: \{houcine.chougrani, steven.kisseleff,  wallace.alvesmartins, symeon.chatzinotas\}@uni.lu.}
\thanks{This research was funded in whole by the Luxembourg National Research Fund (FNR) in the frameworks of the FNR-IPBG INSTRUCT project "INSTRUCT: Integrated Satellite-Terrestrial Systems for Ubiquitous Beyond 5G Communications"  (Grant no. FNR/IPBG19/14016225) and of the FNR-CORE project "MegaLEO: Self-Organised Lower Earth Orbit Mega-Constellations"  (Grant no. C20/IS/14767486). For the purpose of open access, the authors have applied a Creative Commons Attribution 4.0 International (CC BY 4.0) license to any Author Accepted Manuscript version arising from this submission. }
}
\maketitle
\thispagestyle{empty}
\pagestyle{empty}
\begin{abstract}
The satellite component is recognized as a promising solution to complement and extend the coverage of future Internet of things (IoT) terrestrial networks (TNs). In this context, a study item to integrate satellites into narrowband-IoT (NB-IoT) systems has been approved within the 3rd Generation Partnership Project (3GPP) standardization body. However, as NB-IoT systems were   initially conceived for TNs, their basic design principles and operation might require some key  modifications when incorporating the satellite component. These changes in NB-IoT systems, therefore, need to be carefully implemented in order to guarantee a seamless integration of both TN and non-terrestrial network (NTN) for a global coverage. This paper addresses this adaptation for the random access (RA) step in NB-IoT systems, which is in fact the most challenging aspect in the NTN context, for it deals with multi-user time-frequency synchronization and timing advance  for data scheduling. In particular, we propose an RA technique which is robust to typical satellite channel impairments, including long delays, significant Doppler effects, and wide beams,  without requiring any modification  to the current NB-IoT RA waveform.  Performance evaluations demonstrate  the proposal's capability of addressing different NTN configurations recently  defined by 3GPP for the 5G new radio system.
\end{abstract}
\begin{IEEEkeywords}
NB-IoT, 3GPP, Random Access, 5G, NTN, Satellite. 
\end{IEEEkeywords}

\section{Introduction} \label{Introduction}
\IEEEPARstart{I}{n} recent years, a high connectivity demand has started to be experienced in wireless communications.
Virtually everyone and everything need to be connected. 
This universal connectivity poses serious  challenges to terrestrial radio coverage. In fact, the geography of the current uncovered areas is far more diverse than the covered ones, with fjords, mountains, islands, ice, deserts, and vast distances, leading to stringent constraints in terms of backhauling, power availability, site
construction and maintenance~\cite{NGMN}.
In this context, non-terrestrial networks (NTNs) are a promising solution to complement terrestrial networks (TNs) for global coverage extension~\cite{3GPP38913v14}. An NTN refers to a network, or segment of networks,  using radio frequency (RF) resources on board of a satellite or unmanned aerial system (UAS) platform. The primary role of an NTN in this context is to complement the TN services in under-served areas, to improve the TN service reliability, especially for mission-critical services, and to enable the network scalability by means of efficient multicast/broadcast resources for data delivery~\cite{3GPP38913v14,Olt9210567}. 

In 2017, the 3rd Generation Partnership Project (3GPP) started to study the integration of satellites into the 5G ecosystem~\cite{3GPP38811v0}.
The study was completed in 2019 during the corresponding 3GPP November meeting
with the technical reports TR 38.821~\cite{3GPP38821v16} and TR 38.811~\cite{3GPP38811v15}. 
Then, a new normative working item (WI) for 5G new radio (NR) in NTN was approved in 3GPP Release 17~\cite{RP193144thales}.
Recently, a new study WI for narrowband Internet of things (NB-IoT) support in NTN has also been approved in 3GPP Release 17~\cite{RP193235MediaTek}. 


NB-IoT is a recent cellular technology standardized by 3GPP that aims to provide improved coverage for a massive number of low-throughput low-cost devices with low device power consumption in delay-tolerant applications. Prospective applications
include utility metering, environment monitoring, asset tracking,
municipal light, and waste management, to name but a few~\cite{lin2016random}.
However, in many cases these devices are deployed in remote areas which are out of reach of the TN. By integrating NB-IoT into an NTN, 3GPP aims to provide a standardized solution enabling global IoT operation anywhere on Earth. 
In this direction, numerous companies (e.g. Eutelsat and MediaTek) and start-ups (e.g. OQTechnology, Sateliot, and Kepler Communication Inc.) have already started the integration of NB-IoT into NTNs via satellite.
However, the current NB-IoT protocol~\cite{3GPP36211,3GPP36331} was initially designed for TNs and does not account for satellite constraints, such as larger propagation delays, stronger Doppler effects (in case of non-geostationary satellites), limited on-board power availability, etc.

In order to adopt NB-IoT in satellite systems, some modifications to the NB-IoT protocol are therefore required. In this context, and from the physical layer (PHY) perspective, one of the major challenges is the \emph{random access} (RA) which manages the uplink synchronization and requests for scheduling data transmissions. In fact, the current NB-IoT RA preamble namely \emph{Narrowband Physical Random Access Channel} (NPRACH) cannot support the satellite channel impairments (e.g. strong Doppler, long delays) which are more severe compared to those faced by TNs.

The above problem is a new challenge to be faced at PHY level. The most relevant related work in this area is by MediaTek Inc.~\cite{charbit2020space}. The authors have considered a user equipment (UE)  without built-in global navigation satellite system (GNSS) capabilities and proposed a new solution to tackle the relatively large residual frequency offset during the RA reception with relatively mild changes to the existing NB-IoT RA preamble. The proposed solution relies on the introduction of a fractional frequency-hopping sequence per preamble to remove the ambiguity among UEs in presence of large frequency offsets. The authors, however, have not explicitly pointed out the solution for managing the relatively large differential delay that may impact the orthogonality of UEs in the network. Another work in~\cite{kodheli2020random} has analyzed the RA procedure and proposed some system level recommendations, such as reducing the beam width (i.e. cell) in order to handle the large differential delay and residual frequency offset of UEs within this beam using the current preamble, and/or the proposal of a new preamble with longer cyclic prefix (CP) and larger bandwidth to cope with differential delay and frequency offsets. The first recommendation (i.e. small cells) is not straightforward, for it requires a complex antenna technology that may lead to increased size and power consumption of the low Earth orbit (LEO) satellite. The second solution requires considerable changes in the standard, which is something that both 3GPP and stakeholders try to avoid for flexibility, time-to-market, and backward compatibility reasons.

In this work, we address the aforementioned issues without any modification to the current NB-IoT RA preamble. The proposed method consists of the combination of a system level solution that overcomes the problem of a large residual frequency offset with a signal processing solution that tackles the large differential delay. In order to reduce the residual frequency offset, we take advantage of the initial downlink synchronization. More specifically, the UE uses the estimated frequency offset during the downlink to pre-compensate the uplink RA transmission. For the differential delay problem, the inherent $2\pi$-ambiguity problem is solved using a discrimination criterion based on the estimation of Doppler rate.\footnote{The time-varying Doppler shift.} In summary, the main contributions of this paper are:
\begin{enumerate}
    \item A detailed mathematical analysis including all satellite impairments, i.e. delay, frequency offset, and Doppler rate, is presented with an efficient RA receiver design for integrating NB-IoT into NTN via LEO satellites. The designed receiver supports a coverage up to five times the original maximum coverage in TN, and does not require any modifications to the current preamble waveform, which makes it a seamless solution when integrating the satellite into the NB-IoT system.
    \item The Doppler rate corresponding to each UE is estimated within the satellite coverage. This allows for the NTN system to take this information into account for UE data scheduling, which is very useful for the system to preserve uplink multiple access orthogonality. 
    \item PHY-based simulations and performance evaluations in terms of preamble detection probability as well as Doppler and time-of-arrival (ToA) estimation accuracy are provided under realistic scenarios recently defined by 3GPP in the context of NTN.
\end{enumerate}

The remainder of the paper is organized as follows.
In Section~\ref{sec:background}, a background on the 3GPP-based solutions and recommendations to integrate satellites into a 5G system is provided.
Section~\ref{sec:System Model} contains a system overview, including the system parameters, link budget, LEO channel impairments, and the NB-IoT RA preamble. In Section~\ref{sec:TN detection}, the preamble detection and ToA estimation method for TN is briefly described. Then, a novel method for preamble detection and ToA estimation in NTN is proposed in Section~\ref{sec:preamble detect NTN}.
A comprehensive set of simulation results is  provided and discussed in Section~\ref{sec:NumericalResults}. 
The concluding remarks are drawn in  Section~\ref{Conclusion}.

\section{3GPP-based satellite integration into 5G}
\label{sec:background}
NB-IoT is a recent cellular technology standardized by 3GPP that inherits from the existing long term evolution (LTE) technology.
The radio access is based on orthogonal frequency-division multiple access (OFDMA) for downlink and single-carrier frequency-division multiple access (SC-FDMA) for uplink with 180-kHz bandwidth.

\begin{figure}[!t]
\centering
\includegraphics[width = 0.4\textwidth]{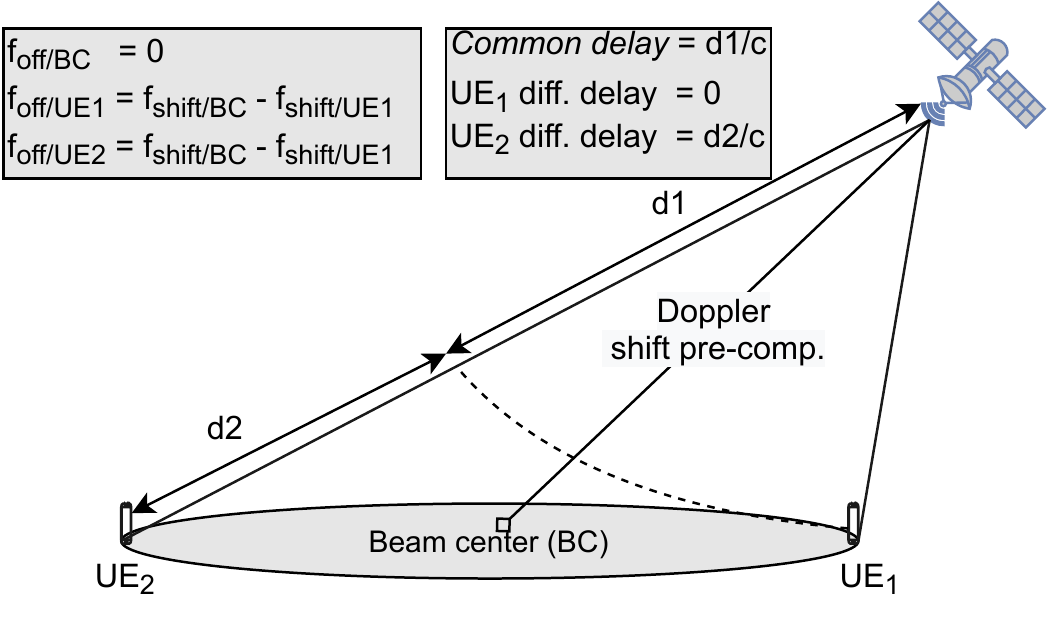}
\caption{Doppler shift pre-compensation mechanism and  common and differential delay illustration for NTN.}
\label{fig:Doppler_delay_solution}
\end{figure}

As we pointed out in the previous section, the adoption of NB-IoT in satellite systems requires some modifications either to the system and/or to the protocols. 
However, since NB-IoT is a 3GPP-based protocol, the recent 3GPP recommendations in~\cite{3GPP38821v16,3GPP38811v15} to integrate 5G-NR into NTN could be adopted for NB-IoT. Those recommendations are based on the type of UE in the NTN system, as follows~\cite{3GPP38821v16,3GPP38811v15}:
\begin{itemize}
    \item \emph{UE with GNSS capability}: the UE performs the pre-compensation of Doppler shift\footnote{Shift of the signals' frequency content due to the motion of either the receiver, the transmitter, or both of them.} and delay propagation. This is possible thanks to the knowledge of satellite \emph{ephemeris} and available UE location. Depending on the prediction accuracy,  additional delay and frequency offsets are assumed to be handled by the protocol~\cite{3GPP38811v15}.
    \item \emph{UE without GNSS capability}: the satellite performs the pre-compensation of Doppler shift at the center of the beam on the ground and broadcasts the \emph{common delay}\footnote{This common delay corresponds to the propagation time between the next generation node B (gNB) and the closest point on Earth in the satellite beam coverage.} to all UEs inside the concerned beam to be taken into account for the uplink transmission, as illustrated in Fig.~\ref{fig:Doppler_delay_solution}. In this case, the additional delay, called \emph{differential delay (DD)} in 3GPP terminology, and frequency offset in the beam (e.g. UE$_{1}$-DD and $f_{\rm off/UE_{1}}$ in Fig.~\ref{fig:Doppler_delay_solution}) are assumed to be handled by the protocol~\cite{RP193144thales}.
\end{itemize}

In this context, one of the major challenges encountered in the integration of 5G-NR into NTN from a PHY perspective is RA.
In 5G-NR as well as in NB-IoT, RA manages the uplink synchronization and requests for scheduling data transmissions.
The uplink synchronization means that the base station (BS) has both to detect (and identify) all active UEs in its coverage area and to estimate their round-trip delays (RTDs). 
Through this, the propagation delay between each UE and the BS is acquired/estimated, which enables attaining a common timing reference. 
The acquired delay allows the BS to perform timing advance needed to keep the orthogonality among multiple UEs, which is typically required in SC-FDMA systems. The estimation of RTD refers to a ToA estimation, whereas the user detection refers to NPRACH preamble detection.
It is worth noting that this operation is essential for a successful system operation.
In fact, RA is the first phase of system operation and covers the first messages from each UE to the BS. Hence, a wrong detection and/or an erroneous ToA estimation would lead to increased system latency, overall performance degradation, and even service outage in extreme cases.

Regarding 5G-NR RA, it was concluded in~\cite{3GPP38821v16} that, in the case of UEs with GNSS capability (assumption for pre-compensation of timing and frequency offset at the UE side based on GNSS and \emph{ephemeris} information), existing RA preambles can be reused if the knowledge of UE's geo-location is available and meets the prescribed accuracy level. It was also stated that additional enhancements, such as  repetitions and/or larger subcarrier spacing (SCS),  are necessary to ensure uplink coverage.
In contrast, for UEs without GNSS capabilities,~\cite{3GPP38821v16} concluded that enhanced/new RA preambles should be proposed.
This paper addresses the same problem but for NB-IoT integration into NTNs considering LEO satellites. More specifically, the paper provides a solution to integrate the NB-IoT NPRACH into the NTN context.
We focus here on UEs without GNSS capabilities, since the applicability of GNSS solution for NB-IoT in NTN is questionable. In fact, the GNSS solution poses additional challenges, such as: 
\begin{itemize}
    \item unclear update procedure for the satellite \emph{ephemeris} at the UE level.
    \item reduced battery lifetime for the UEs due to the frequent estimation of Doppler and delay along with their compensation, thus requiring substantial additional signal processing.
\end{itemize}
Furthermore, 3GPP targets both cases, i.e. UE with and without GNSS capability, which eventually means that the GNSS solution should not be viewed as the key strategy.

\section{System model} \label{sec:System Model}
In this section, we describe the  system model adopted in this work. The description includes the system architecture, the link budget, the satellite channel impairments, and the NPRACH preamble. {For clarity's sake, Table \ref{tab:System model symbols} summarizes the mathematical symbols used in this section.}
\begin{table}[t]
\caption{{Mathematical symbols used in the system model}}
	\label{tab:System model symbols}
	\centering
	\begin{tabular}{|l|p{0.29 \textwidth}|}
		\hline
				\textbf{\textit{Symbol}} & \textbf{\textit{Definition}} \\
		\hline\hline
		         $G/T$   & Antenna-gain-to-noise-temperature \\
		        EIRP  & Effective isotopic radiated power\\
		       $k_{\rm B}$         &  Boltzmann constant\\
		        $PL_{\rm FS}$ & Free space propagation loss\\
		      ${\rm PL}_{\rm A}$ & Atmospheric gas losses \\      
		      ${\rm PL}_{\rm S}$  & Shadowing margin\\
			${\rm PL}_{\rm AD}$	 & Additional
losses due to the scintillation phenomena \\
			 ${\rm BW}_{\text{NB-IoT}}$   & NB-IoT channel bandwidth\\
				${\rm BW}_{\mathrm{NPRACH}}$  & NPRACH bandwidth\\
				$g$      & Ratio between NB-IoT channel and NPRACH bandwidths\\
			CNR      & Carrier-to-noise ratio \\
				SNR & Signal-to-noise ratio\\
				\textit{T}$_{\rm CP}$  & CP length in ms\\
				\textit{L} & Number of identical symbols within one symbol group\\
				\textit{T}$_{\rm SEQ}$ & Length of identical symbols within one symbol group in ms\\
				\textit{P} & Number of symbol groups in one preamble basic unit\\
			$\Delta f$	& Subcarrier spacing between NB-IoT tones/SGs\\
			$N_{\rm rep}$	& Number of preamble basic unit repetition\\
		\hline		   				
		\end{tabular}
\end{table}

\subsection{System architecture}
In the 3GPP framework, six reference scenarios have been specified depending on the orbit nature---whether geosynchronous Earth orbit (GEO) or LEO---, payload type---whether transparent or regenerative---, and beam type---whether steerable or fixed~\cite{3GPP38821v16,3GPP38811v15}. We consider in this work the scenario referred to as \emph{D2} in~\cite{3GPP38821v16} among the NTN reference scenarios from 3GPP.
TABLE~\ref{Beam_layout} provides the satellite parameters for this scenario.
Note that 3GPP has defined two sets of beam layout and RF parameters of the payload, namely: Set-1 and Set-2. Both parameter sets are taken into account in this work. 
The satellite is placed at 600-km altitude (LEO) with a regenerative payload, which  means that the satellite implements fully or partially the functionalities of the classical base station (i.e. gNB). The system is operating in S-band at a carrier frequency of 2 GHz.
The satellite generates several spot-beams over its coverage area to increase its capacity.
These spot-beams are moving along with the satellite. Each spot-beam size, however,  depends on the elevation angle and the satellite configuration, whether it is Set-1 or Set-2 (depending on the 3-dB beam width angular value). 
Moreover, antenna-gain-to-noise-temperature ($G/T$) in Set-1 is greater than in Set-2, which comes as a byproduct of the reduction of the equivalent satellite antenna aperture. For more details  related to the payload configuration and beam layout aspect, we refer the reader to~\cite{ESA9149179, R11907481ESA,3GPP38821v16}, where a thorough discussion can be found.

\begin{table}[t]
\caption{Satellite parameters for UL transmissions}
	\label{Beam_layout}
	\centering
		\begin{tabular}{|P{0.18 \textwidth}|P{0.11 \textwidth}|P{0.11\textwidth}|}
		\hline
				\textbf{Parameters} & \textbf{Set-1} & \textbf{Set-2}\\
		\hline		
			 Altitude & 600 km & 600 km \\
		\hline
		        Payload & Regenerative & Regenerative\\
		 \hline
		        Operating band   & S-band (2~GHz) & S-band (2~GHz)\\
		 \hline
                Moving beams & Yes & Yes \\
		 \hline
		        Satellite antenna aperture & 2 m & 1 m \\
		 \hline 
		        $G/T$  & 1.1 dB$\cdot$K$^{-1}$ & -4.9~dB$\cdot$K$^{-1}$\\
		 \hline
		        3 dB beam width & $\approx$ 4.4127$^{\circ}$ &  $\approx$ 8.8320$^{\circ}$\\
		 \hline
	   				
		\end{tabular}
\end{table}

\subsection{Link budget}\label{subsec:link_budget}
In this subsection, we calculate the link budget of the aforementioned system architecture for the RA in UL transmissions. We consider a minimum elevation angle of 30 degrees, which represents a typical elevation angle in satellite-based systems~\cite{3GPP38811v15}. The general formula for the link budget is derived from~\cite{R1-1913224ESA}. It accounts for all the gains and losses in the propagation medium from transmitter to receiver. Accordingly, the carrier-to-noise ratio (CNR) can be computed as follows:
\begin{align}
\mathrm{CNR[dB]}=&\mathrm{EIRP[dBW]} + G/T\mathrm{[dB]} - k_{\rm B}{\rm [dBW/K/Hz]}\nonumber\\
&-{\rm PL}_{\rm FS}{\rm [dB]} - {\rm PL}_{\rm A}{\rm [dB]} - {\rm PL}_{\rm S}{\rm [dB]}\nonumber\\
&-{\rm PL}_{\rm AD}{\rm [dB]} - 10\log_{10}({\rm BW}_{\text{NB-IoT}}{\rm [Hz]}),
\label{eq:linkbudget}
\end{align}
where $\mathrm{EIRP}$ is the effective isotopic radiated power of the transmitting antenna, $k_{\rm B} = -228.6$~dBW/K/Hz is the Boltzmann constant, ${\rm PL}_{\rm FS}$ represents the free space propagation loss, ${\rm PL}_{\rm A}$ corresponds to atmospheric gas losses, ${\rm PL}_{\rm S}$ is a shadowing margin, ${\rm PL}_{\rm AD}$ denotes some underlying additional losses due to the scintillation phenomena, and ${\rm BW}_{\text{NB-IoT}}$ is the NB-IoT channel bandwidth.
TABLE~\ref{link_budget} illustrates the link budget for both payload configurations (i.e. Set-1 and Set-2) in the UL. 
The values in the table are similar to the ones suggested in~\cite{3GPP38821v16,R1-1913224ESA}. 
To obtain the signal-to-noise ratio (SNR), the NPRACH bandwidth (i.e. ${\rm BW}_{\mathrm{NPRACH}}$) of 3.75~kHz is considered. This leads to a gain factor of $g = 48$ (i.e. $\frac{{\rm BW}_{\text{NB-IoT}}}{{\rm BW}_{\mathrm{NPRACH}}}$).
Following the recommentation in~\cite{3GPP38821v16}, 6-dB degradation is introduced in the SNR calculation as an additional margin. Hence, the SNR is given by:
\begin{align}
\mathrm{SNR[dB]} = {\rm CNR[dB]} + 10\log_{10}(g) - 6.
\label{eq:SNR}
\end{align}

\begin{table}[t]
\caption{Link budget for NPRACH}
	\label{link_budget}
	\centering
		\begin{tabular}{|p{0.16 \textwidth}|c|c|c|c|}
		\hline
				\textbf{Transmission mode} & \multicolumn{2}{c|}{\textbf{UL Set-1}} & \multicolumn{2}{c|}{\textbf{UL Set-2}} \\
				
		\hline
		        UE elevation angle        & 30$^{\circ}$ & 90$^{\circ}$ &30$^{\circ}$ & 90$^{\circ}$ \\
		\hline
		        Frequency [GHz]  & 2 & 2 & 2 & 2\\
		 \hline
		        TX EIRP [dBm]    & 23.01 & 23.01 & 23.01 & 23.01 \\
		 \hline
		        RX $G/T$ [dB/K]    & 1.10 & 1.10 & -4.9 & -4.9 \\
		 \hline
		        BW$_\text{NB-IoT}$ [kHz]      & 180 & 180 & 180 & 180 \\
		 \hline
		    Free space path loss [dB] &	159.10	& 154.03 &	159.10	& 154.03\\
		 \hline
		 Atmospheric loss [dB]	& 0.07 &	0.07 & 0.07 &	0.07\\
		 \hline
		Shadowing margin [dB]	& 3.00 &	3.00 & 3.00 &	3.00\\
		 \hline
		Scintillation loss [dB]	& 2.20	& 2.20 & 2.20	& 2.20\\
		 \hline
		 Polarization loss [dB] & 0.00	& 0.00 & 0.00	& 0.00 \\
		 \hline
		 Additional losses [dB] & 0.00	& 0.00 & 0.00	& 0.00 \\
		 \hline
		 CNR [dB]               & 5.79 & 10.85 & -0.21 & 4.86 \\
		 \hline
		 BW$_{\rm NPRACH}$ [kHz] & 3.75 & 3.75 & 3.75 & 3.75\\
		 \hline
		 Additional margin [dB] & -6 & -6 & -6 & -6\\
		 \hline\hline
		 \textbf{SNR [dB]}   & \textbf{16.60} & \textbf{21.66} & \textbf{10.60} & \textbf{15.67}\\
		 \hline
	   				
		\end{tabular}
\end{table}

\subsection{Satellite channel impairments}
\label{subsec:NTN_channel}
\begin{figure}[t]
\centering
\includegraphics[width = 0.5\textwidth]{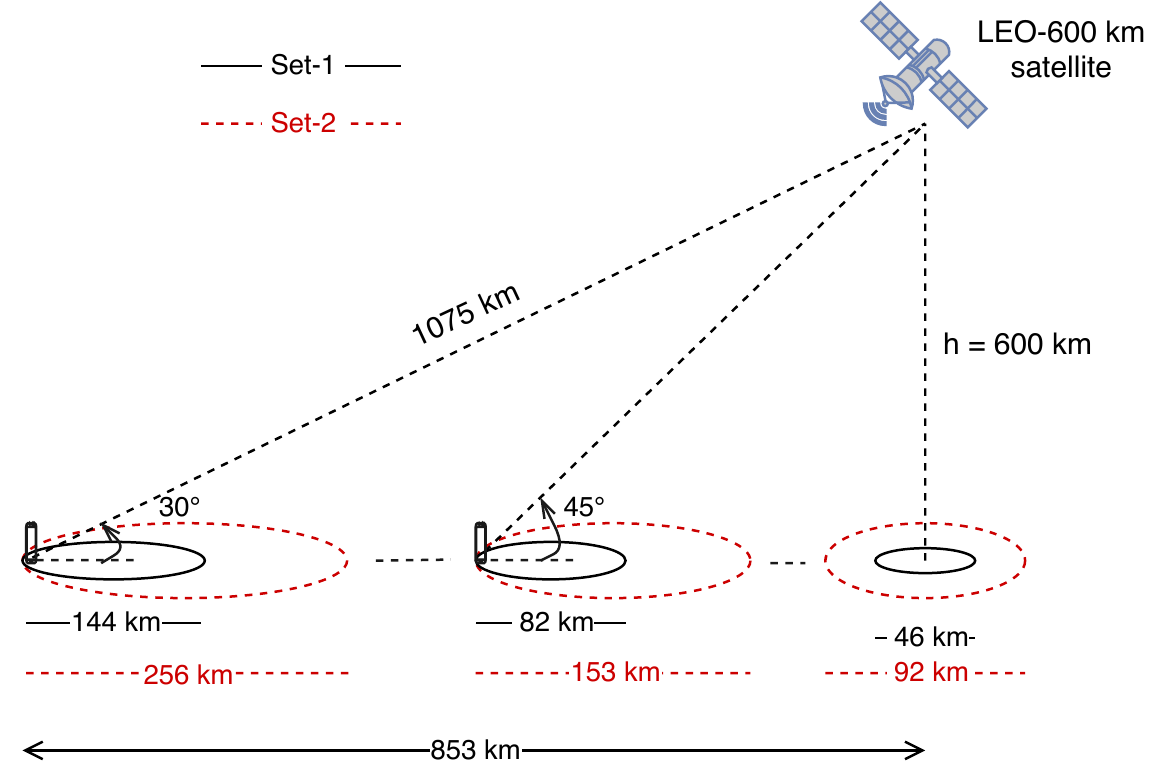}
\caption{Beam layout for LEO satellite at 600 km.}
\label{fig:beamwidth}
\end{figure}
\subsubsection{Long propagation delay}
\label{subsec:sat_delay}
In satellite systems, the propagation delays are long when compared to TN's delays due to much longer distances between the UE and the satellite. In fact, the propagation delay depends on the satellite orbit, type of payload, and elevation angle. In our scenario (i.e. LEO at 600 km with regenerative payload), the minimum and maximum RTDs at 30$^{\circ}$ and 90$^{\circ}$ are  7.17~ms and 4~ms, respectively. Note that the RTD is approximated as twice the propagation delay between the satellite and the UE.
Adopting the recommendations of~\cite{3GPP38821v16} regarding the RTD---i.e., broadcasting the common RTD (C-RTD) to all UEs within each beam---, the maximum differential RTD (MD-RTD) will depend on the beam size, which itself depends on the elevation angle, as shown in Fig.~\ref{fig:beamwidth}.
TABLE~\ref{tab:diff_RTD} provides the MD-RTD values at different elevation angles (after taking into account the C-RTD) for both types of payload configurations, viz. Set-1 and Set-2. 
At higher elevation angles, the beam is small (e.g. 46-km and 92-km beam diameters at 90$^{\circ}$ for Set-1 and Set-2, respectively), whereas at lower elevation the beam becomes larger (e.g. 144-km and 257-km beam diameters at 30$^{\circ}$ for Set-1 and Set-2, respectively).
For Set-1 configuration, the MD-RTD varies from 3.25~$\mu$s to 807.13~$\mu$s, whereas for Set-2 configuration, it starts from 13.04~$\mu$s and reaches up to 1.40~ms.
As it will be discussed later, this is a problematic situation for the RA in NB-IoT.

\begin{table}[t]
\caption{MD-RTD after common delay compensation}
	\label{tab:diff_RTD}
	\centering
		\begin{tabular}{|c|P{0.07 \textwidth}|P{0.07 \textwidth}|P{0.07 \textwidth}|P{0.07 \textwidth}|}
		\hline
				\multirow{4}{{0.065 \textwidth}}{\textbf{Elevation}} & \multicolumn{2}{c|}{\textbf{Set-1}} & \multicolumn{2}{c|}{\textbf{Set-2}}\\
				\cline{2-5}
			& Beam diameter [km] & MD-RTD [$\mu$s] & Beam diameter [km] & MD-RTD [$\mu$s]\\
		\hline
		        30$^{\circ}$ & 144.02 & 807.13 & 256.91 & 1397.23\\
		 \hline
		        45$^{\circ}$ & 82.10  & 369.53  & 153.63 & 658.85\\
		 \hline
                90$^{\circ}$    & 46.23 & 3.25 & 92.70 & 13.04\\
		 \hline
	   				
		\end{tabular}
\end{table}

\subsubsection{Significant Doppler effects}
\label{subsubsec:Large Doppler shift}
\begin{figure}[t]
\centering
\includegraphics[width = 0.5\textwidth]{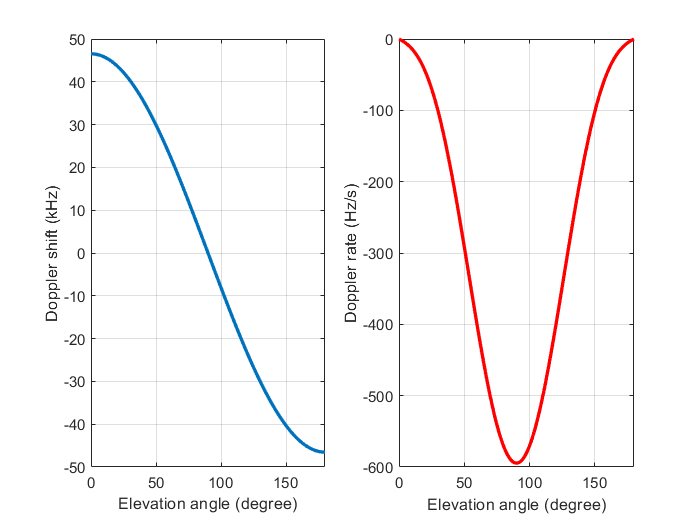}
\caption{Doppler variation depending on the elevation angle.}
\label{fig:Doppler_curves}
\end{figure}
The high speed motion of LEO satellites introduces significant Doppler shifts on the NB-IoT signal as compared to those expected in TN systems.
In our scenario, up to $\pm 41$-kHz Doppler shift (see Fig.~\ref{fig:Doppler_curves}) is observed (with 30$^\circ$ minimum elevation angle).
Furthermore, the received signals are impacted by a significant Doppler rate (see Fig.~\ref{fig:Doppler_curves}) that varies between $-101$~Hz/s (for 30$^{\circ}$ elevation) and $-594$~Hz/s (for 90$^{\circ}$ elevation), thus leading to additional degradation of the detection performance in case of longer preambles/data packets.
Using the Doppler shift pre-compensation technique as defined in~\cite{3GPP38821v16}, the maximum residual frequency offsets (normalized by the carrier frequency) at the satellite receiver for RA are shown in TABLE~\ref{tab:freq.offset}. This maximum residual frequency offset is twice the offset of one-way propagation from satellite to UE.
Note that in NB-IoT, the UE motion is negligible, and the residual frequency offsets account only for satellite motion and the precision of local oscillators (LOs).

\begin{table}[t]
\caption{Residual frequency offset after Doppler pre-compensation}
	\label{tab:freq.offset}
	\centering
		\begin{tabular}{|c|P{0.06 \textwidth}|P{0.08 \textwidth}|P{0.06 \textwidth}|P{0.08 \textwidth}|}
		\hline
				\multirow{6}{{0.065 \textwidth}}{\textbf{Elevation}} & \multicolumn{2}{c|}{\textbf{Set-1}} & \multicolumn{2}{c|}{\textbf{Set-2}}\\
				\cline{2-5}
			& Beam diameter [km] & Max. residual freq. offsets [ppm] & Beam diameter [km] & Max. residual freq. offsets [ppm]\\
		\hline
		        30$^{\circ}$ & 144.02 & $\pm$ 1.28  & 256.91 & $\pm$ 2.68\\
		 \hline
		        45$^{\circ}$    & 58.15 & $\pm$ 1.54 & 153.63 & $\pm$ 3.24 \\
		 \hline
                90$^{\circ}$    & 46.23 & $\pm$ 1.94 & 92.70 & $\pm$ 3.88 \\
		 \hline
	   				
		\end{tabular}
\end{table}

\subsection{NB-IoT random access}
\label{subsec:RA preamble}
\begin{table}[t]
\caption{NPRACH preamble formats for FDD mode}
	\label{tab:preamble_formats}
	\centering
		\begin{tabular}{|P{0.06 \textwidth}|P{0.06 \textwidth}|P{0.024 \textwidth}|P{0.024 \textwidth}|P{0.068 \textwidth}|P{0.058 \textwidth}|}
		\hline
		\rowcolor{Gainsboro!60}
				 \makecell{\textbf{format}} & \makecell{\textbf{$\Delta f$}} & \makecell{\textbf{\textit{P}}} & \makecell{\textbf{\textit{L}}} & \makecell{\textbf{\textit{T}$_{\rm CP}$}} & \makecell{\textbf{\textit{T}$_{\rm SEQ}$}}\\
		\hline		
			 0 & 3.75 kHz & 4 & 5 & 66.67~$\mu$s & 1.33~ms\\
		\hline
		        1 & 3.75 kHz & 4 & 5 & 266.67~$\mu$s & 1.33~ms\\
		 \hline
		        2 & 1.25 kHz & 6 & 3 & 800~$\mu$s & 2.4~ms\\
		 \hline
	   				
		\end{tabular}
\end{table}

\begin{figure*}[h!]
\centering
\includegraphics[width = 0.9\textwidth]{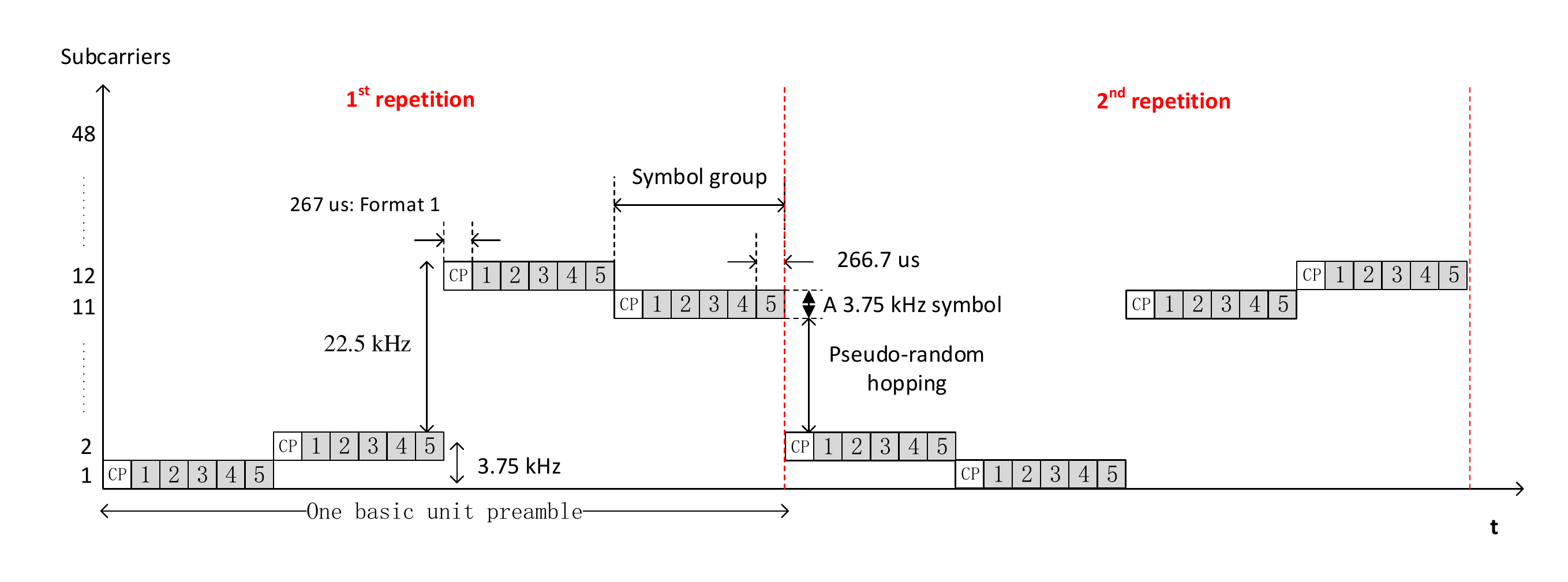}
\caption{NPRACH preamble format 1 structure.}
\label{fig:NPRACH_Preamble_format1}
\end{figure*}

The RA preamble in NB-IoT, known as NPRACH preamble, was originally proposed in~\cite{ericsson2015,ericsson2015x2,ericsson2016,ericsson2016x2} and then adopted by 3GPP and integrated into the NB-IoT Release~13~\cite{3GPP36300}, followed by an enhancement in Release~15 ~\cite{3GPP36211v15,3GPP36331}. 
In the latter, a new preamble format is introduced for range enhancement and time-division multiplexing (TDD). Note that, in Releases 13 and 14, only frequency-division multiplexing (FDD) mode was supported. It is worth noting here that our work focuses only on FDD since it is the preferable mode in NTN, as described in~\cite{3GPP38821v16}. This is because TDD mode requires a guard time that directly depends on the propagation delay between UE and satellite to prevent UE from transmitting and receiving simultaneously. However, such a guard time might be excessive in NTN and would lead to a very inefficient radio interface.

In FDD mode, there are three NPRACH formats, namely: format 0, format 1, and format 2, as shown in TABLE~\ref{tab:preamble_formats}. 
The NPRACH formats with different CP, symbol group sizes, and symbol group repetitions have been designed according to the targeted cell sizes.
The preamble consists of a set of \emph{symbol groups} (SGs).
An SG consists of a CP of length $T_{\rm CP}$ and a sequence of $L$ identical symbols with total length $T_{\rm SEQ}$.
In the 3GPP standard, $P$ SGs are treated as the basic unit of the preamble.
The basic unit can be repeated up to $N_{\rm rep} = 2^j, j\in \{0, 1, \ldots, 7\}$ times for coverage extension.
Accordingly, the length of a preamble equals $P\cdot N_{\rm rep}$ SGs.
The NPRACH transmission supports either a 3.75-kHz or a 1.25-kHz subcarrier spacing (SCS, i.e. $\Delta f$).

The hopping pattern is fixed within the basic unit of $P$ SGs.
Symbol groups in preamble format 0 and 1 (with 3.75-kHz SCS) hop by one or six subcarriers in frequency, whereas symbol groups in format 2 (with 1.25-kHz SCS) hop by one, three, or eighteen subcarriers in frequency~\cite{3GPP36211v15}.
Fig.~\ref{fig:NPRACH_Preamble_format1} shows an example of preamble format 1 with two preamble basic unit repetitions (i.e. $N_{\rm rep} = 2$).
Note that when repetitions are configured, the hopping between the basic units is no longer fixed, but it follows a pseudo-random selection procedure defined in~\cite{3GPP36211v15}.

Currently, there are three possible CP lengths.  
CP lengths of 66.7~$\mu$s (for format 0), 266.67~$\mu$s (for format 1), and 800~$\mu$s (for format 2) are designed to support cell radius of up to 10~km, 40~km, and 120~km,  respectively. In terms of propagation delay, the three formats account for maximum RTDs of 66.67~$\mu$s, 266.67~$\mu$s, and 800$\mu$s, respectively.

\section{Preamble detection and ToA estimation in TN}
\label{sec:TN detection}
In this section, we provide a brief review of our recently proposed method~\cite{Houcine2020} for NPRACH detection and ToA estimation in TN. This is an essential part to understand the proposed solution in this paper. Note that in NB-IoT TN system, the Doppler is negligible due to very low mobility conditions.
For clarity’s sake, we employ the same notation and the same variable definitions as provided in the 3GPP standard~\cite{3GPP36211v15} and in our previous work~\cite{Houcine2020} in the rest of this paper. {These mathematical notations are summarized in Appendix \ref{Appendix: Mathematical symbols}.}

\subsection{Signal model}
\label{subsec:signal model TN}
Based on~\cite{Houcine2020}, the transmitted NPRACH baseband signal can be written as
\begin{equation} 
s_{m,i}[n] = \sum_{k}{S_{m,i}[k]} {\rm e}^{{\rm j}2\pi\frac{k}{N}n},
\end{equation}
where $s_{m,i}[n]$ is the $n$-th sample of the time-domain waveform of $i$-th symbol in the $m$-th SG, whereas $S_{m,i}[k]$ denotes the $i$-th symbol on the $k$-th subcarrier during the $m$-th SG; in this case,  $S_{m,i}[k] = 1$ for all $m,~i$. 
Furthermore, the sample index $n$ belongs to the set $\{N_{m,i}-N_{\rm CP},...,N_{m,i}+N-1\}$, in which $i \in \{0,...,L-1\}$, with $L$ being the number of symbols in one SG, and $N_{m,i} = mN_{g} + iN$, with $N_{g}= N_{\rm CP}+LN$ being the size of one SG, $N_{\rm CP}$ denoting the CP size, and $N$ denoting the size of a symbol.

It was shown in~\cite{Houcine2020}, that the received signal after OFDM demodulation (i.e. CP removal and discrete Fourier transform, DFT,  application) assuming a negligible inter-carrier interference (ICI) can be expressed as
\begin{align}\label{generalrxsignal2}
{Y_{m,i}} = \,& h_{m}{\rm e}^{{\rm j}2\pi f_{\rm off}( mN_{g} + iN-D)}\nonumber\\
&\times {\rm e}^{-{\rm j}2\pi n_{\rm SC}^{\rm RA}(m)\frac{D}{N}}\left(\frac{1-{\rm e}^{{\rm j}2\pi f_{\rm off}N}}{1-{\rm e}^{{\rm j}2\pi f_{\rm off}}}\right)+W_{m,i}\,,
\end{align}
where $f_{\rm off}$ is the carrier frequency offset (CFO) normalized by the sampling frequency, $D$ is the RTD normalized by the symbol duration, $h_{m}$ is the channel coefficient for the $m$-th SG, $n_{\rm SC}^{\rm RA}(m)$ is the subcarrier occupied by the $m$-th SG, and $W_{m,i}$ is the noise term.
Combining the symbols within the same $m$-th SG, we get the so-called SG-sum (SG-S) as follows:
\begin{eqnarray}\label{eq:SG-S}
Y_{m}\hspace*{-2mm}&=&\hspace*{-2mm}\sum_{i=0}^{L-1}{{{Y_{m,i}}}}=h_{m}{\rm e}^{{\rm j}2\pi f_{\rm off}( mN_{g} -D)}\notag\\
\hspace*{-2mm}&\times&\hspace*{-2mm}{\rm e}^{-{\rm j}2\pi n_{\rm SC}^{\rm RA}(m)\frac{D}{N}}\left(\frac{1-{\rm e}^{{\rm j}2\pi f_{\rm off}LN}}{1-{\rm e}^{{\rm j}2\pi f_{\rm off}}}\right)~{+~W_{m}}{.}
\end{eqnarray}

\subsection{Preamble detection and ToA estimation}
\label{subsec:ToA in TN}
For the clarity of exposition, the noise term and some constant factors are omitted here. For more details, we refer the reader to~\cite{Houcine2020}.
The process to detect the preamble and to estimate the ToA is as follows:
\begin{enumerate}
    \item Perform differential processing  by multiplying the $m$-th SG-S with the complex conjugated $(m+1)$-th SG-S. Defining $\Delta(m) = n_{SC}^{RA}(m+1) - n_{\rm SC}^{\rm RA}(m)$ as the hopping step between the $m$-th and $(m+1)$-th SGs, the differential processing gives 
     \begin{eqnarray}
     \label{generalrxsignal4}
         Z_{m,1}\hspace*{-2mm}&=&\hspace*{-2mm}{Y_{m}Y_{m+1}^{*}} \propto {\rm e}^{-{\rm j}2\pi f_{\rm off}N_{g}}{\rm e}^{{\rm j}2\pi \Delta(m) \frac{D}{N}}
     \end{eqnarray}
     \item Construct an array $v$ such that the position of $Z_{m,1}$ in $v$ corresponds to the hopping step $\Delta(m)$. If there are multiple $Z_{m,1}$ with equal value of $\Delta(m)$, i.e. $\Delta(m_{1}) = \Delta(m_{2})$ with $m_{1} \neq m_{2}$, then their values are first summed up before being inserted in $v$.
     \item Perform one dimensional DFT (1D-DFT) with $N_{\rm DFT}$ points on $v[n]$ to get
    \begin{equation}\label{eq:ToAFFT}
        \begin{split}
        V[k] &= \sum_{n=0}^{N_{\rm DFT}-1}{v[n]}{\rm e}^{-{\rm j}2\pi k\frac{n}{N_{\rm DFT}}},
        \end{split}
    \end{equation}
     \item Combine the results non-coherently over $N_{\rm RX}$ receive antennas to get
    \begin{equation}\label{eq:antenna_combin}
        \begin{split}
            X[k] &= \sum_{N_{\rm RX}}\lvert{V[k]}\rvert^{2}.
        \end{split}
    \end{equation}
      \item Determine  $k_{\max} = {\rm argmax}_{k} \{X[k]\}$ and define $X_{\max} = X[k_{\max}]$.
     \item Compare $X_{\max}$ to a predefined threshold (cf.~\cite{Houcine2020}). If $X_{\max}$ is greater than or equal to the threshold the preamble presence is declared, otherwise it is considered absent.
     \item If the preamble is declared as present, the ToA is estimated as
    \begin{equation}\label{eq:calculate_ToA}
        \hat{D} = \frac{k_{\max}}{N_{\rm DFT}\Delta f}.
    \end{equation}
\end{enumerate}

Note that the most interesting property of the above method is its insensitivity to CFO when estimating ToA. This is due to the perfect elimination of the CFO effect in step 4). Indeed, the term ${\rm e}^{-{\rm j}2\pi f_{\rm off}N_{g}}$ is common for all $Z_{m,1}$ symbols (i.e. common for all $v[n]$) and affects only the phase of $v[n]$, not its magnitude. Similarly, since the DFT operation is linear, this factor affects only the phase of $V[k]$, not its magnitude, such that by taking the absolute of $V[k]$, the impact of CFO is completely eliminated~\cite{Houcine2020}.

\section{Preamble detection and ToA estimation in NTN}
\label{sec:preamble detect NTN}

In this section, we first provide the solution to handle high frequency offsets observed in NTN-LEO-based systems. Then, a detailed mathematical analysis is provided before describing the solution for preamble detection and ToA estimation. {Before describing the proposed method, we notice here that in NTN systems, UEs may experience multiple visible satellites, leading to inter-beam interference. Although this issue concerns the whole NTN operation system including the actual data transmissions (not only the preamble detection), we believe that this issue can be tackled using the classical color schemes, usually employed in practical satellite communication where different carriers are configured for the adjacent satellites/beams. Through this the inter-beam interference will be reduced to a minimum, which would have a negligible impact on the performance of the proposed method.}

\subsection{Solution for residual frequency offset}
\label{subsec:solution FO}
\begin{figure}[t!]
\centering
\includegraphics[width = 0.49\textwidth]{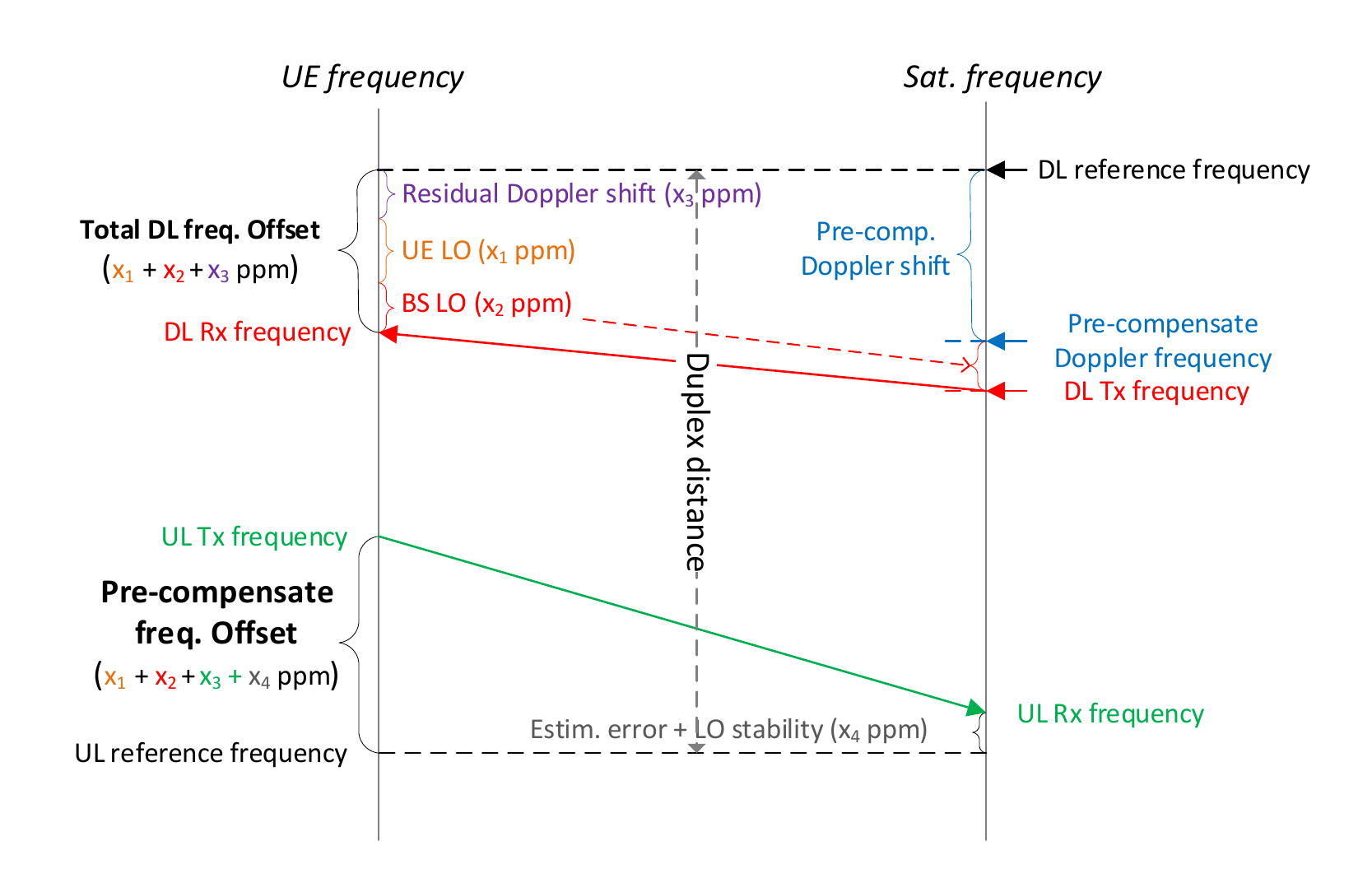}
\caption{Residual CFO reduction scheme.}
\label{fig:freq_precorrect}
\end{figure}

We described in Section \ref{sec:background} the 3GPP solution to reduce large Doppler shifts due to LEO satellites' motions. However, the residual frequency offset seen by the satellite (in uplink) is still high and it can even exceed the largest SCS in NPRACH, i.e. 3.75~kHz (see TABLE~\ref{tab:freq.offset}). This frequency misalignment causes ICI over the subcarriers and also a mismatch between the sampling frequencies used at transmission and reception.
To tackle this issue, we propose a system level solution to reduce the overall frequency offsets seen by the satellite to a value that could be handled by the RA preamble. Accordingly, in addition to the Doppler pre-compensation at the satellite, we propose to pre-compensate the residual frequency offset at the UE level using the initial downlink frequency synchronization (known as \emph{initial cell-search procedure}). Fig.~\ref{fig:freq_precorrect} illustrates the different steps of the proposed system level solution. The general idea is that the UE estimates the frequency offset in the downlink synchronization process. Hence, it can shift the central frequency to the opposite side of the \emph{estimated frequency offset} for uplink transmission. 

The process is done as follows:
\begin{itemize}
    \item UE acquires a downlink frequency synchronization during initial cell-search procedure, which is not perfectly accurate. In practice, the frequency offset seen by the UE at this level consists of 3 components. The first one is the residual Doppler shift that depends on the UE location within the beam. The second and third components are frequency offsets introduced by imperfect accuracy of both satellite and UE LOs.
    \item UE tracks frequency after the initial downlink synchronization due to the high Doppler rate (e.g. during the demodulation of narrowband physical broadcast channel).
    \item UE updates the initial estimated frequency offset considering the tracking information just before the UL transmission.
    \item UE pre-compensates the calculated frequency offset in the previous step and sends its RA signal. 
\end{itemize}

At the end, the residual frequency offset seen by the satellite receiver will depend on the performance of downlink frequency synchronization, frequency tracking, and LO stability (LO frequency drift in time). To the best of our knowledge, there is no work providing the downlink synchronization performance for NB-IoT under a LEO satellite scenario. However, if we can consider the downlink synchronization results for 5G-NR reported recently in~\cite{R1-1908049Huawei}, the residual frequency offset after downlink frequency synchronization was about 600 Hz (at 99\% of cases).
In such an NTN scenario, note that preamble format 2 is not suitable because of its small SCS (i.e. only 1.25 kHz) which makes it too sensitive to a frequency offset. A CFO of 600 Hz represents about half of its SCS. For this reason, we adopt in this work the preamble format 1 since it has the largest SCS (i.e. 3.75 kHz) associated with the longest CP (i.e. 267~$\mu$s). However, if future NB-IoT UEs guarantees lower frequency offsets, preamble format 2 can be adopted with our proposed method. {It is worth pointing out that the UE hardware does not require any modification, if the Doppler pre-compensation can be done at the satellite as assumed in our method, the maximum CFO seen at the UE level will be comparable to the terrestrial scenario and, therefore, resolvable by typical NB-IoT UE implementation.} 

\subsection{Solution for Differential RTD}
\label{subsec:solution RTD}

Regarding the differential RTD (D-RTD), one can see that the current NPRACH preambles (see TABLE~\ref{tab:preamble_formats}) cannot address beams at lower elevation, since the timing uncertainty (up to 1.4~ms) is much larger than the current CP length (266.67~$\mu$s for format 1).
In order to better understand the proposed solution to support long D-RTDs, the reasons behind the design of NPRACH preamble are listed below:
\begin{itemize}
    \item A CP length needs to be at least equal to the expected maximum RTD.
    \item The frequency hopping is used to facilitate the ToA (i.e. RTD) estimation at the BS.
    \item The maximum ToA that can be estimated corresponds to the reciprocal minimum hopping step (i.e. ToA $\in \{0,\dots, \frac{1}{\Delta(m)_{\min}}\}$).
\end{itemize}
\begin{figure}[t!]
\centering
\includegraphics[width = 0.49\textwidth]{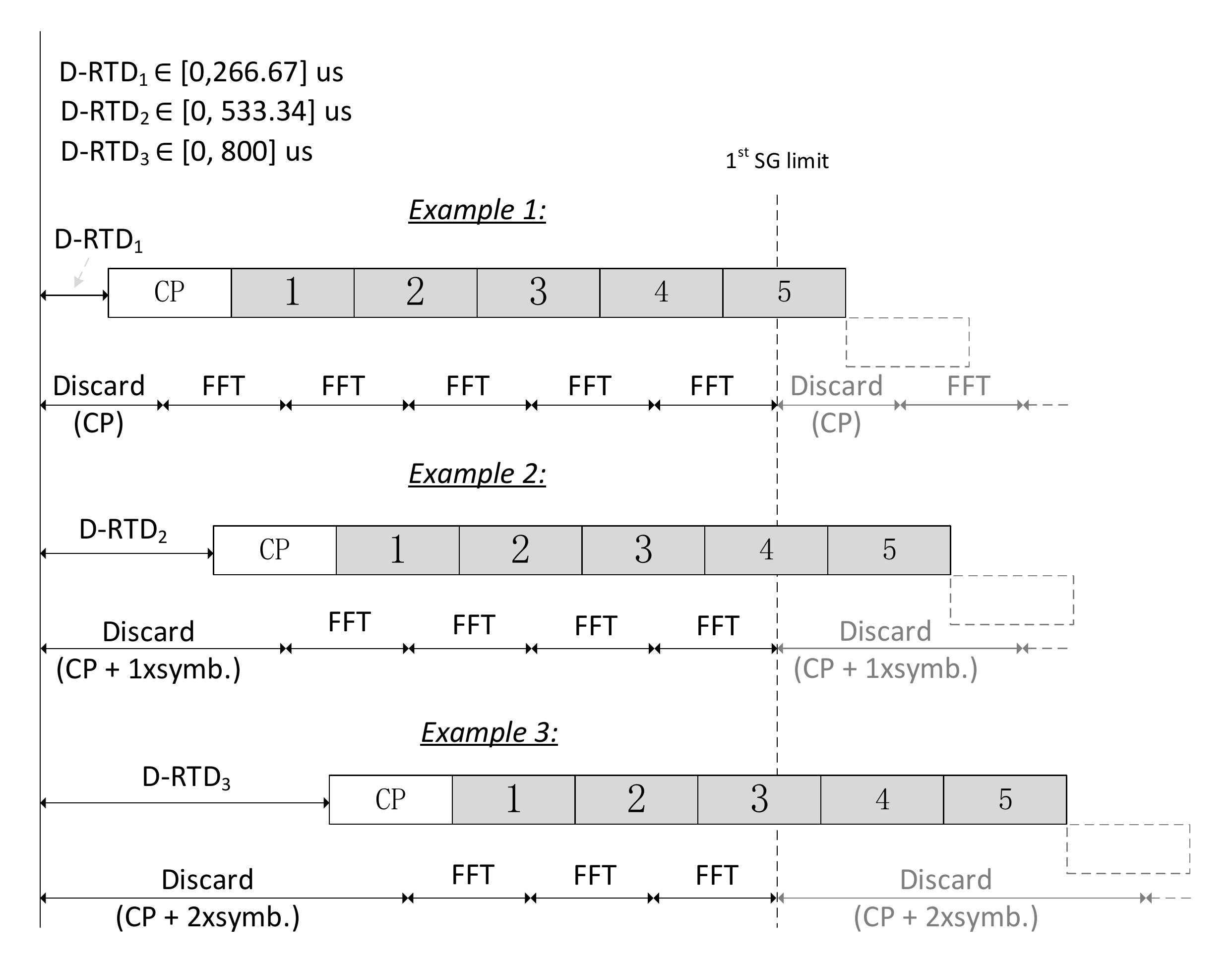}
\caption{OFDM demodulation examples with the proposed CP extension scheme.}
\label{fig:CP extension}
\end{figure}
Regarding the CP, we propose to extend its length by considering the current one (with 266.67~$\mu$s) plus additional symbols from the SG. The number of additional symbols depends on the MD-RTD (i.e. beam width). Note that this is possible since the current CP consists of multiple repetitions of the same symbol. { Fig. \ref{fig:CP extension} shows, from the receiver perspective, three examples representing three scenarios where the D-RTD belongs to three different coverage levels with MD-RTD of 266.67 $\mu$s, 533.34 $\mu$s and 800 $\mu$s, respectively. For example, if the MD-RTD is 800~$\mu$s (Example 3 in Fig. \ref{fig:CP extension}), the receiver considers the new CP, namely CP$_{\rm new}$, to be the current one plus 2 symbols from the sequence within the SG. Accordingly, the receiver will discard (i.e. remove) in this case the CP$_{\rm new}$ when the OFDM demodulation is done.}
Interestingly, the aforementioned processing does not imply any change neither on the transmitter nor on the standard. Hence, it does not limit the practicality of the proposed solution.
\begin{figure}[t!]
\centering
\includegraphics[width = 0.5\textwidth]{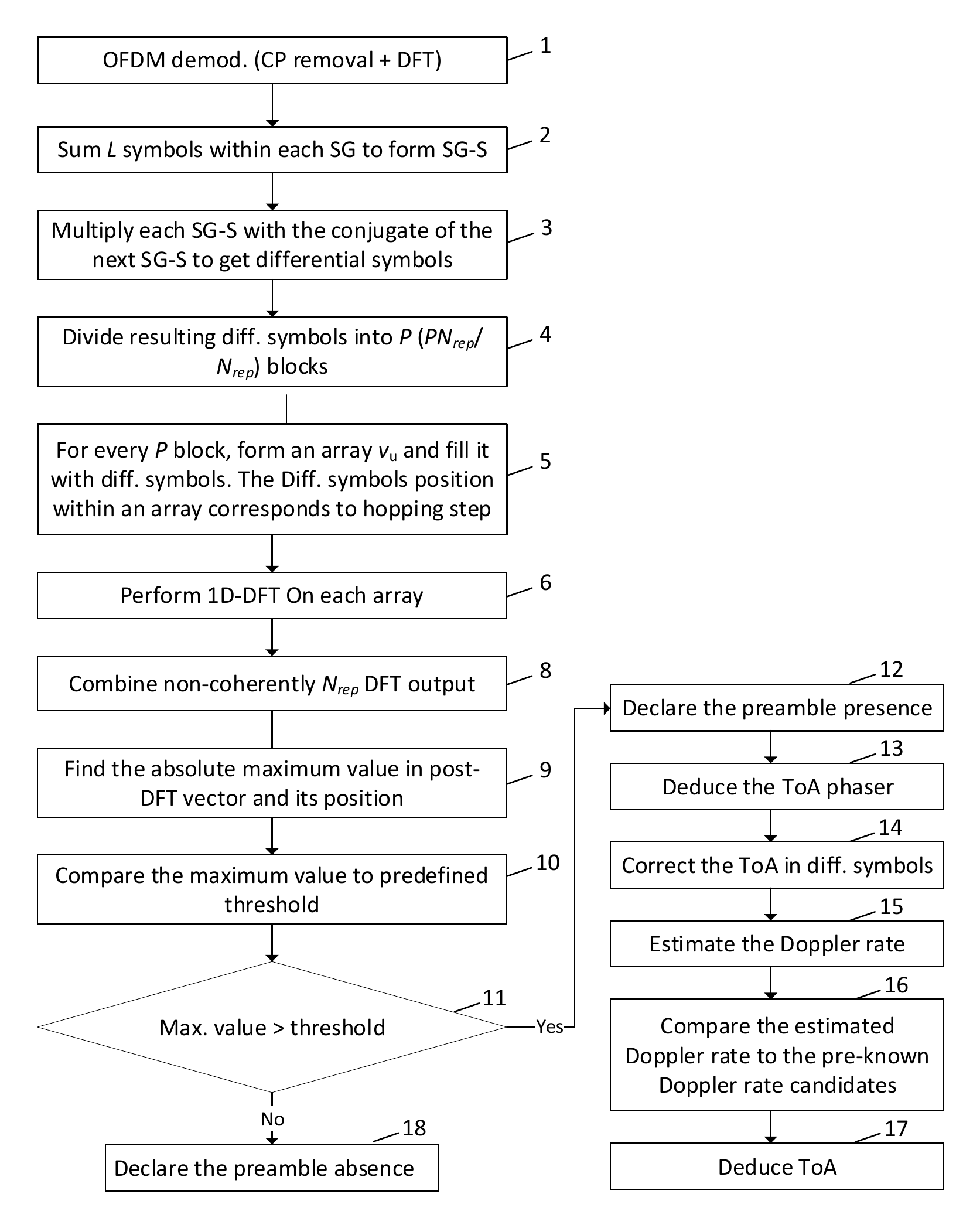}
\caption{Flowchart of the overall solution for preamble detection and ToA estimation in NTN.}
\label{fig:flowchart_ToA_NTN}
\end{figure}

The other issue that we need to solve, is the estimation of ToA larger than 266.67~$\mu$s (in case of using the preamble format 1), which represents the maximum ToA that can be estimated according to the minimum hopping distance of 3.75~kHz.
A flowchart illustrating the overall solution for ToA estimation as well as the preamble detection is depicted in Fig.~\ref{fig:flowchart_ToA_NTN}. The flowchart aims to provide a roadmap for the reader to refer and follow the different steps detailed in the forthcoming description. The step numbers that will refer to in the description correspond to the numbers in the flowchart in Fig.~\ref{fig:flowchart_ToA_NTN},  unless otherwise specified.

First, let us take a look at the received signal in the presence of Doppler rate. Omitting the noise term (for the clarity of exposition), the received signal can be reformulated as
\begin{eqnarray}
y_{m,i}[n]\hspace*{-2mm}&=&\hspace*{-2mm}h_{m} {\rm e}^{{\rm j}2\pi[ f_{\rm off}(n-D) + \frac{1}{2}\alpha (n-D)^{2}]} s_{m,i}[n-D] \notag\\
\hspace*{-2mm}&=&\hspace*{-2mm} h_{m} {\rm e}^{{\rm j}2\pi[ f_{\rm off}(n-D) + \frac{1}{2}\alpha (n-D)^{2}]}\notag\\
\hspace*{-2mm}&\times&\hspace*{-2mm} \sum_{k} {\rm e}^{{\rm j}2\pi\frac{k}{N}(n-D)},
\label{eq:Rx_signal}
\end{eqnarray}
where $\alpha$ is the Doppler rate normalized by the squared sampling frequency. 
By removing the CP (i.e. $N_{\rm CP}$ samples) and performing DFT (Step 1), we obtain
\begin{eqnarray}
Y_{m,i}[l] \hspace*{-2mm}&=&\hspace*{-2mm} \sum_{n=N_{m,i}}^{N_{m,i}+N-1}{y_{m,i}[n]{\rm e}^{-{\rm j}2\pi ln/N}}\notag\\
\hspace*{-2mm}&=&\hspace*{-2mm}\sum_{n= N_{m,i}}^{N_{m,i}+N-1}h_{m} {\rm e}^{{\rm j}2\pi [ f_{\rm off}(n-D) + \frac{1}{2}\alpha (n-D)^{2}]}\notag\\
\hspace*{-2mm}&\times&\hspace*{-2mm}\sum_{k}{
{\rm e}^{{\rm j}2\pi \frac{k-l}{N}n} {\rm e}^{-{\rm j}2\pi \frac{k}{N}D}}.
\label{eq:post_FFT_Rx_signal}
\end{eqnarray}

Assuming (for the analytical derivation) that the energy leakage from other subcarriers is small, and when $l= n_{\rm SC}^{\rm RA}(m)$, \eqref{eq:post_FFT_Rx_signal} can be expressed as
\begin{eqnarray}
Y_{m,i} \hspace*{-2mm}&=&\hspace*{-2mm} h_{m}{\rm e}^{{\rm j}2\pi f_{\rm off}(N_{m,i}-D)} {\rm e}^{{\rm j}\pi \alpha(N_{m,i}^{2}+D^{2}-2N_{m,i}D)}\notag\\
\hspace*{-2mm}&\times&\hspace*{-2mm}
{\rm e}^{-{\rm j}2\pi n_{\rm SC}^{\rm RA}(m)\frac{D}{N}}
\sum_{n^\prime = 0}^{N-1}{\rm e}^{{\rm j}2\pi f_{\rm off}n^\prime} \notag\\ \hspace*{-2mm}&\times&\hspace*{-2mm}{\rm e}^{{\rm j}\pi \alpha(n^{\prime2} + 2n^\prime(N_{m,i}-D))},
\label{eq:detailed_post_FFT_Rx_signal}
\end{eqnarray}
where $n^\prime = n-N_{m,i}$.

Similarly to~\cite{morelli2002doppler, mengali2013synchronization}, since $\alpha\ll1$, the approximation ${\rm e}^{{\rm j}\pi \alpha(n^{\prime2}-2n^\prime D)} \approx 1$
can be made in \eqref{eq:detailed_post_FFT_Rx_signal} for $0\leq n^\prime \leq N-1$. Correspondingly, we obtain
\begin{eqnarray}
Y_{m,i} \hspace*{-2mm}&=&\hspace*{-2mm} h_{m}{\rm e}^{{\rm j}2\pi f_{\rm off}(N_{m,i}-D)} {\rm e}^{{\rm j}\pi \alpha(N_{m,i}^{2}+D^{2}-2N_{m,i}D)}\notag\\
\hspace*{-2mm}&\times&\hspace*{-2mm}
{\rm e}^{-{\rm j}2\pi n_{\rm SC}^{\rm RA}(m) \frac{D}{N}}\left(
\frac{1-{\rm e}^{{\rm j}2\pi (f_{\rm off} + \alpha N_{m,i})N}}{1-{\rm e}^{{\rm j}2\pi (f_{\rm off} + \alpha N_{m,i})}}\right).
\label{eq:Rx_signal_with_assumpt1}
\end{eqnarray}

By combining the signals within the same $m$-th SG, we get the SG-S (Step 2)
\begin{eqnarray}
Y_{m} \hspace*{-2mm}&=&\hspace*{-2mm} \sum_{i=0}^{L^\prime-1}{{{Y_{m,i}}}} =  h_{m}{\rm e}^{{\rm j}2\pi f_{\rm off}(mN_{g} -D)} \notag\\
\hspace*{-2mm}&\times&\hspace*{-2mm}  {\rm e}^{{\rm j}\pi \alpha [(mN_{g})^{2} + D^{2} - 2mN_{g}D]} {\rm e}^{-{\rm j}2\pi n_{\rm SC}^{\rm RA}(m) \frac{D}{N}}\notag\\
\hspace*{-2mm}&\times&\hspace*{-2mm}\sum_{i = 0}^{L^\prime-1}{\rm e}^{{\rm j}2\pi f_{\rm off}iN}{{\rm e}^{{\rm j}\pi \alpha [(iN)^{2} +  2mN_{g}iN-2iND]}}  \notag\\
\hspace*{-2mm}&\times&\hspace*{-2mm} \frac{1-{\rm e}^{{\rm j}2\pi (f_{\rm off} + \alpha (mN_{g}+iN))N}}{1-{\rm e}^{{\rm j}2\pi (f_{\rm off} + \alpha (mN_{g}+iN))}},
\label{eq:combine within Ymi}
\end{eqnarray}
where $L^\prime = L-L_{\rm CP}$, with $L_{\rm CP}$ denoting the number of symbols employed to extend the CP within the SG.

Reasoning as with \eqref{eq:Rx_signal_with_assumpt1} (i.e. $\alpha\ll1$), approximations\footnote{ Approximations in \eqref{eq:Rx_signal_with_assumpt1} and \eqref{eq:combine within Ymi} are possible only for indexes representing a short time duration. With indexes representing a long time duration (e.g. index $m$) those approximations are no longer reasonable.} ${\rm e}^{{\rm j}\pi\alpha[(iN)^{2} -2iND]} \approx 1$, ${\rm e}^{{\rm j}2\pi\alpha iN^{2}} \approx 1$ and ${\rm e}^{{\rm j}2\pi\alpha iN} \approx 1$ can be made in \eqref{eq:combine within Ymi} for $0\leq i \leq L^\prime-1$. This leads to
\begin{eqnarray}
Y_{m} \hspace*{-2mm}&=&\hspace*{-2mm} h_{m}{\rm e}^{{\rm j}2\pi f_{\rm off}(mN_{g} -D)} {\rm e}^{-{\rm j}2\pi n_{\rm SC}^{\rm RA}(m) \frac{D}{N}}\notag\\
\hspace*{-2mm}&\times&\hspace*{-2mm} {\rm e}^{{\rm j}\pi \alpha [(mN_{g})^{2} + D^{2} - 2mN_{g}D]}\notag\\
 \hspace*{-2mm}&\times&\hspace*{-2mm} \frac{1-{\rm e}^{{\rm j}2\pi (f_{\rm off} + \alpha mN_{g})L^\prime N}}{1-{\rm e}^{{\rm j}2\pi (f_{\rm off} + \alpha mN_{g})}}\,.
\label{eq:eq:Rx_signal_final}
\end{eqnarray}

The preamble detection and ToA estimation method in NTN is performed in the same way as described in Section~\ref{sec:TN detection} for TN systems. However, one modification may be included if the Doppler rate is too high. This is because the assumption of having a constant CFO during the entire NPRACH period no longer holds. Hence, when constructing the array $v$ with $Z_{m,1}$ symbols (Step 2 in the TN method), we cannot sum up multiple $Z_{m,1}$ with equal value of $\Delta(m)$ (particularly for far apart symbols), since they do not have the same phase due to the variant frequency offset. Furthermore, to eliminate the CFO impact on the ToA estimation, the factor ${\rm e}^{-{\rm j}2\pi f_{\rm off}N_{g}}$ needs to be a common factor for all $Z_{m,1}$, which is not the case with time-varying CFO.
Nevertheless, for the analytical derivation, we assume that the CFO variation within one preamble basic unit\footnote{The preamble basic unit contains $P$ SGs, where $P$ can be set according to TABLE~\ref{tab:preamble_formats}.} is negligible. Accordingly, the proposed method for the NTN can be outlined as follows:
\begin{itemize}
    \item Perform differential processing by multiplying the $m$-th SG-S with the complex conjugated $(m+1)$-th SG-S (Step 3). This gives: 
    \begin{eqnarray}
    \label{eq:Zm_withdrift}
        \hspace*{-3mm}Z_{m,1}\hspace*{-2mm}&=&\hspace*{-2mm}{Y_{m}Y_{m+1}^{*}} \propto {\rm e}^{-{\rm j}2\pi f_{\rm off}N_{g}}{\rm e}^{{\rm j}2\pi \Delta(m) \frac{D}{N}}\notag\\
        \hspace*{-3mm}\hspace*{-2mm}&\times&\hspace*{-2mm} {\rm e}^{-{\rm j}\pi \alpha (N_{g}^{2} + 2mN_{g}^2 - 2N_{g}D)} {\rm e}^{-{\rm j}\pi \alpha N_{g}(L^\prime N-1)}.
    \end{eqnarray}
    \item For each preamble basic unit,\footnote{There are $N_{\rm rep}$ basic units in a preamble.} we construct an array $v_{u}$ with the $Z_{m,1}$ symbols belonging to the basic unit $u$. Their position in $v_{u}$ corresponds to their hopping step $\Delta(m)$ (Steps 4 and 5). Thus, we end up with $N_{\rm rep}$ arrays. 
    \item Perform 1D-DFT with $N_{\rm DFT}$ points for each $v_{u}[n]$ to get (Step 6)
    \begin{equation}\label{eq:ToAFFT_NTN}
        \begin{split}
        V_{u}[k] &= \sum_{n=0}^{N_{\rm DFT}-1}{v_{u}[n]}{\rm e}^{-{\rm j}2\pi k\frac{n}{N_{\rm DFT}}},
        \end{split}
    \end{equation}
    \item Combine non-coherently over $N_{\rm rep}$ and $N_{\rm RX}$ receive antennas to get (Step 7)
        \begin{equation}\label{eq:anetnna_combin_NTN}
            \begin{split}
                        X[k] &= \sum_{u=0}^{N_{\rm rep}-1}\sum_{N_{\rm RX}}\lvert{V_{u}[k]}\rvert^{2}.
            \end{split}
        \end{equation}
    \item Perform Steps 5 and 6 of the TN method (Steps 8 to 12 in Fig.~\ref{fig:flowchart_ToA_NTN}).
\end{itemize}

\subsection{Correct ToA selection}
\label{subsec:ToA selection}
\subsubsection{ToA selection concept}
\label{subsec:ToA selection concept}
As mentioned earlier, we need to solve the problem of estimating ToAs larger than the maximum allowed by the NPRACH preamble (i.e. $\mathrm{ToA} > \frac{1}{\Delta(m)_{\min}}$).
In the proposed method, the estimated ToA will always be in the range $[0,\frac{1}{\Delta(m)_{\min}}]$. But, if we analyze the ToA term in \eqref{eq:Zm_withdrift}, i.e. ${\rm e}^{{\rm j}2\pi \Delta(m) \frac{D}{N}}$, we  notice that there may be  a 2$\pi$-phase-ambiguity problem.
In other words, the phase rotation caused by D-RTD is prone to a 2$\pi$-ambiguity by multiples of $\frac{1}{\Delta(m)_{\min}}$ in the time estimates. Given that $\Delta(m)_{\min} =$ 3.75~kHz (for format 1), the time domain ambiguity is in multiples of 266.67~$\mu$s. This means that all possible D-RTDs will be estimated as D-RTD modulo 266.67~$\mu$s. For example, D-RTDs of 66.67~$\mu$s, 333.34~$\mu$s and 600~$\mu$s will all have the same phase rotation (i.e. $2\pi \Delta(m) \frac{D}{N}$) and a ToA estimate of 66.67~$\mu$s. Note that depending on the CP length, a list of ToA candidates can be obtained. In principle, if the new CP is equal to $c\times 266.67~\mu$s, we will end up with $c$-ToA candidates. To select the right one, we propose a discrimination criteria based on the estimation of the underlying Doppler rate.

It is well-known that the Doppler rate depends on the position of the UE within the beam, similarly to the D-RTD. Also, the satellite knows the Doppler rate and D-RTD of each geographical location within its coverage area (thanks to \emph{ephemeris} and beam layout knowledge). Hence, the estimation of Doppler rate during the NPRACH reception allows us to select the correct ToA (Step 16) by comparing the estimated Doppler rate with the available one predicted by the satellite. More explicitly, the estimated Doppler rate will be compared to all Doppler rates corresponding to all D-RTD candidates (Step 15). To illustrate this strategy, consider the following example. In a given beam, assume that the MD-RTD is 800~$\mu$s and the estimated ToA is 66.67~$\mu$s. Due to the phase ambiguity there will be three D-RTD candidates, namely 66.67~$\mu$s, 333.34~$\mu$s, and 600~$\mu$s with corresponding Doppler rate (e.g. Set-2 beam with minimum elevation angle of 42.1$^{\circ}$) of -305~Hz/s, -261~Hz/s, and -225~Hz/s, respectively. Next, assume that the estimated Doppler rate is -275~Hz/s. Since -275~Hz/s is closer to -261~Hz/s than  the other two Doppler rate candidates, we can conclude that the most likely ToA is 333.34~$\mu$s.

Note that in this solution, the Doppler rate estimation does not require a very high accuracy, especially at higher elevation angles. In practice, the estimation error needs only to be less than half of the minimum difference between any two ToA candidates. At lower elevation angles, where wider beams  are expected  (i.e. higher D-RTD), Doppler rates are small and the number of ToA candidates is high. In this situation, a very good Doppler rate estimation accuracy is required, since the Doppler rate separating the ToA candidates is small. Nonetheless, for higher elevation angles, where high Doppler rates and low ToA candidates are expected, the accuracy is relaxed due to larger Doppler rate separation between the ToA candidates. In addition, note that the Set-1 configuration requires better accuracy compared to Set-2 configuration, since it has a narrower beam.
In the former example, the error should be smaller than 18~Hz/s.

\subsubsection{Doppler rate estimation}
After we outlined the concept of selecting the correct ToA candidate based on the Doppler rate estimation, we describe in the following how the Doppler rate estimator is designed. Note that the Doppler rate is always negative and has a maximum value at 90$^\circ$ elevation angle.

We start by compensating the estimated ToA in \eqref{eq:Zm_withdrift} as follows (Step 13):
\begin{eqnarray}
\label{eq:tm}
    t_{m}\hspace*{-2mm}&=&\hspace*{-2mm} Z_{m,1}{\rm e}^{-{\rm j}2\pi \Delta(m) \frac{\hat{D}}{N}}\notag\\
    \hspace*{-2mm}&\propto&\hspace*{-2mm} {\rm e}^{-{\rm j}2\pi f_{\rm off}N_{g}}{\rm e}^{-{\rm j}\pi \alpha (N_{g}^{2} + 2mN_{g}^2 - 2N_{g}D)}\notag\\
    \hspace*{-2mm}&\times&\hspace*{-2mm} {\rm e}^{-{\rm j}\pi \alpha N_{g}(LN-1)}. 
\end{eqnarray}

Interestingly, one can notice that all the terms in \eqref{eq:tm} are constants except the term ${\rm e}^{-{\rm j}2\pi \alpha N_{g}^2 m}$ that depends on $m$. Similarly to the ToA estimator, the Doppler rate $\alpha$ can be estimated using an approximate maximum-likelihood (ML) estimator based on 1D-DFT with $N_{\rm DFT}$ points. In this case, we obtain
\begin{eqnarray}
\label{eq:IFFT_drift}
    T[p] &= \sum_{n=0}^{N_{\rm DFT}-1}{t[n]}{\rm e}^{-{\rm j}2\pi p\frac{n}{N_{\rm DFT}}}\,.
\end{eqnarray}

In case of multiple receive antennas, the combination of the signals is done non-coherently by summing up squared magnitudes of $T[p]$ over $N_{\rm RX}$ antennas, i.e. $J[p] = \sum_{N_{\rm RX}}\lvert{T[p]}\rvert^{2}$. 
Note that by taking the absolute square of $T[p]$, we eliminate the impact of the CFO given by ${\rm e}^{-{\rm j}2\pi f_{\rm off}N_{g}}$ as well as the other constant terms, since they affect only the phase of $T[p]$ and not its magnitude.

Let $p_{\max}$ be the index of the maximum value of $J[p]$. The Doppler rate can be estimated as 
\begin{eqnarray}
\label{eq:drift_estim}
     \hat{\alpha} &= -\frac{p_{\max}}{N_{\rm DFT}N_{g}^{2}}.
\end{eqnarray}

However, one can observe from \eqref{eq:tm} that $w_{\alpha} = 2\pi \alpha N_{g}^{2}$ in $t_{m}$ symbols is very small, since $\alpha \ll 1$. Hence, to be able to estimate $\alpha$ by the above method (i.e. 1D-DFT) with a good estimation accuracy, a very large number of DFT points is needed. For example, if a resolution 
of 6 Hz/s is desirable,  $N_{\rm DFT} = \frac{1}{6\times N_{g}^{2}} \approx 2^{16}$ is required. This leads to a very high computational complexity, namely $\mathcal{O}\left(N_{\rm DFT}\log_{2}(N_{\rm DFT})\right)\approx 2^{20}$ (considering radix-2 fast Fourier transform implementations), which may become a burden for the system operation, especially due to the limited computational resources and power available on-board of the satellite.

To reduce this computational complexity and enable an efficient implementation of the method, we consider the approximate ML estimator via a numerical evaluation of the discrete-time Fourier transform (DTFT). The key idea herein is to restrict the search space in frequency domain only to the possible Doppler rate values (depending on the scenario) rather than the whole spectrum. More specifically, note that the maximum Doppler rate that can be estimated with the above method corresponds to $\alpha_{\max} = -\frac{1}{N_g^{2}} = - 390.625$~kHz/s, whereas the maximum value in the adopted scenario is only $\alpha_{\max}^{\rm sc} = -594$~Hz/s, which occupies a very small part of the considered spectrum. In addition, note that the minimum Doppler rate is $\alpha_{\min}^{\rm sc} = -101$~Hz/s (see Section~\ref{subsec:NTN_channel}). Furthermore, each beam, depending on its size, has a minimum and maximum Doppler frequency drift between [$\alpha_{\min}^{\rm sc}, \alpha_{\max}^{\rm sc}$] depending on the elevation angle. All these aspects  reduce further the space search in the spectrum, which can be exploited in order to considerably reduce the computational complexity and increase the estimation accuracy.

At first, the DTFT of $t_m$ with $N_{\rm DTFT}$ points is calculated: 
\begin{eqnarray}
\label{eq:IFFT_drift}
    T({\rm e}^{{\rm j}w_{q}}) &= \sum_{n=0}^{N_t-1}{t[n]}{\rm e}^{-{\rm j}w_{q}n},
\end{eqnarray}
where $N_{t} = PN_{\rm rep}-1$ is the number of symbols in $t[n]$, and $w_{q} =- 2\pi\delta_{\alpha}q$ with $ \delta_{\alpha} = \frac{(\alpha_{\max}^{\rm sc}-\alpha_{\min}^{\rm sc})N_{g}^{2}}{N_{\rm DTFT}}$ is the desired resolution controlled by the number of DTFT points $N_{\rm DTFT}$, and $q \in \left\{\frac{\alpha_{\min}^{\rm sc}}{\alpha_{\max}^{\rm sc} -\alpha_{\min}^{\rm sc}},...,\frac{\alpha_{\max}^{\rm sc}}{\alpha_{\max}^{\rm sc} -\alpha_{\min}^{\rm sc}}\right\}\cdot N_{\rm DTFT}$.

Then, similarly to the 1D-DFT based method, we combine non-coherently over $N_{\rm RX}$ antennas:
 \begin{eqnarray}
 \label{eq:antenna_combin_drift}
     J[q] = \sum_{N_{\rm RX}}\lvert{T({\rm e}^{{\rm j}w_{q}})}\rvert^{2}.
 \end{eqnarray}

Let $q_{\max}\in [1,N_{\rm DTFT}]$ be the index of the maximum value of $J[q]$. Then the Doppler rate is given by (Step 14):
\begin{eqnarray}
\label{eq:drift_estim}
     \hat{\alpha} &= \frac{q_{\max}\delta_{\alpha}}{N_{g}^{2}} + \alpha_{\min}^{\rm sc}.
\end{eqnarray}

The complexity of the DTFT-based approximate ML estimator will be ${\cal O}(N_{\rm DTFT}\times N_{t})$. 
To show the gain of the proposed ML estimator based on DTFT compared to the standard DFT (implemented via FFT), we consider the same example as before with a target resolution of 6 Hz/s. For the largest spotbeam from Set-2 configuration at 30$^{\circ}$ (see TABLE~\ref{tab:diff_RTD}), the maximum and minimum Doppler rate will be -206 Hz/s and -101 Hz/s, respectively. In this case, we need to have $N_{\rm DTFT} = \frac{\alpha_{\max}^{\rm sc}-\alpha_{\min}^{\rm sc}}{-6} \approx 18$. Assuming a preamble with $N_{\rm rep} = 64$ (i.e. $N_{t} = 255$), the foreseen  complexity will be only ${\cal O}(N_{\rm DTFT}\times N_{t}) \approx 4600$. Thus, we reduce the complexity by a factor of $\frac{2^{20}}{4600}\approx 228$, leading to a substantial improvement in term of computational complexity.

Ultimately, based on the estimate of the Doppler rate, it is possible to discriminate the candidates of ToA (Step 15) and resolve the corresponding ambiguity (Step 16), as described in Section~\ref{subsec:ToA selection concept}.

\section{Simulation Results and discussion}
\label{sec:NumericalResults}
\begin{table}[t]
\caption{Simulation parameters}
	\label{tab:Simulation param}
	\centering
		\begin{tabular}{|l|p{0.29 \textwidth}|}
		\hline
				\textbf{Parameter} & \textbf{Value} \\
		\hline\hline
		Antenna config. {\cite{3GPP36104}} & 1 Tx, 2 Rx \\
		{No. of active UEs}           & 1 \& 3 \\
		Channel  & AWGN\\
		Preamble format        &  1 \\
		Subcarrier spacing    & 3.75 kHz\\
		No. of repetitions  & 32 and 64\\
		Min. Elevation & 30$^{\circ}$ (for Set-1) and 31$^{\circ}$ (for Set-2)\\
		MD-RTD & 533.3~$\mu$s, 800~$\mu$s, 1066.7~$\mu$s, and 1333.4~$\mu$s \\
		Timing uncertainty         & Uniformly drawn from [0, MD-RTD]~$\mu$s \\
		Residual Freq. offset      &  Uniformly drawn from [-600, 600] Hz\\
		$N_{\rm DFT}$ (ToA estim.)             & 256 points \\
		$N_{\rm DTFT}$ ($\alpha$ estim.)    & 62 points\\
		\hline		   				
		\end{tabular}
\end{table}

In this section, we provide simulation results to assess the performance of the proposed method. The PHY-based simulation parameters are summarized in TABLE~\ref{tab:Simulation param}. 
The antenna configuration is set to 1 transmitter antenna (at the UE) and 2 receiver antennas (at the satellite) similarly to the TN system.
One active UE is considered first to obtain an upper-bound performance for the proposed method. Then, three active UEs with adjacent subcarriers are simulated to represent a realistic scenario specifically 
 accounting for ICI effects \cite{Ericsson160277}.
{Regarding the channel model, it is worth noting here that the NTN propagation environment can exhibit NLOS/ multipath effects. 3GPP study related to the channel model in TR 38.811 \cite{3GPP38811v15} has developed a channel model for 5G-NR that supports a range of deployment including urban, suburban and rural scenarios. However, in contrast to terrestrial propagation environment, it was highlighted in TR 38.811 \cite{3GPP38811v15} that the large distance in NTN leads to lower delay spread and thus to higher coherence bandwidth. In fact, TR 38.811 \cite{3GPP38811v15} defined 100 ns to be the maximum Delay spread for our considered scenario (i.e. LEO at 600 km). This corresponds in the worst case (i.e. performance lower bound) to 200 kHz of coherence bandwidth. Hence, the NB-IoT NTN channel is considered to be flat due to its narrow band nature (i.e. 180 kHz), especially for the random access (i.e. 45 kHz considering maximum hopping of 12 subcarriers). Consequently, and as stated in TR 38.811 \cite{3GPP38811v15}, the link level simulation under flat fading does not require any channel modeling, only additive white Gaussian noise (AWGN) is considered.}
Both 32 and 64 repetitions ($N_{\rm rep}$) are configured. The considered minimum elevation angles for Set-1 and Set-2 payload configurations are 30${^\circ}$ and 31${^\circ}$, respectively. This latter represents the MD-RTD that the proposed method can address in case of Set-2 configuration. The MD-RTDs for our scenario that exceed the typical maximum RTD in the TN system (i.e. 266.67~$\mu$s) are considered to show the coverage extension provided by the proposed method. Accordingly, the timing uncertainty is uniformly chosen from [0, MD-RTD]~$\mu$s. The maximum residual frequency offset is chosen to be 600~Hz according to our discussion in Section~\ref{subsec:solution FO}. Finally, 256 points DFT (implemented via FFT) and 62 points DTFT are set for ToA and Doppler rate estimation, respectively. 
\subsection{3GPP performance metrics}
3GPP has not yet defined performance metrics for NB-IoT in NTN. However, we adopt in this work the same 3GPP performance metrics as in TN. Accordingly, the performance metrics are the probability of preamble detection and false alarm. More specifically, the probability of preamble detection should be equal to $99\%$ (i.e. missed detection rate below $1\%$), and false alarm probability should be less than or equal to $0.1\%$.
According to~\cite{3GPP36104}, the probability of detection is defined as the conditional probability of correct detection of the preamble when the signal is present. There are several error cases: 
\begin{itemize}
    \item detection of a wrong preamble (different than the one sent);
    \item no detection of any preamble;
    \item correct preamble detection but with the wrong timing (ToA) estimation.
\end{itemize}

This latter occurs if the ToA estimation error is larger than $3.646~\mu$s.

The false alarm probability is defined as the conditional total probability of erroneous detection of the preamble when the input contains only noise, i.e. in absence of the useful signal.
\begin{figure*}[!ht]
  \centering
  \subfloat[NTN beam with MD-RTD of 1333.4~$\mu$s ]{\includegraphics[width=0.45\textwidth]{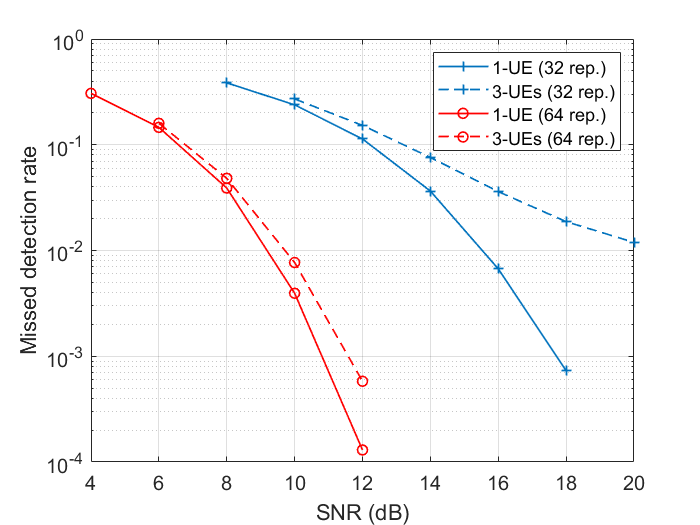}\label{fig:max_RTD_1}}
  \hfill
  \subfloat[NTN beam with MD-RTD of 1066.7~$\mu$s]{\includegraphics[width=0.45\textwidth]{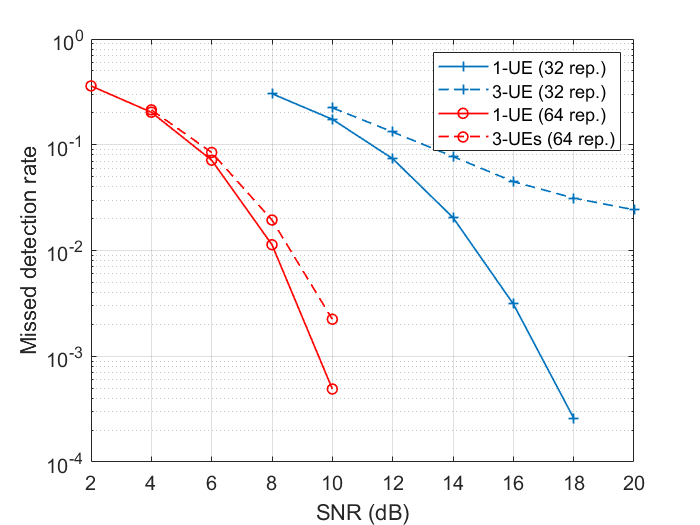}\label{fig:max_RTD_2}}
  \\
  \subfloat[NTN beam with MD-RTD of 800~$\mu$s ]{\includegraphics[width=0.45\textwidth]{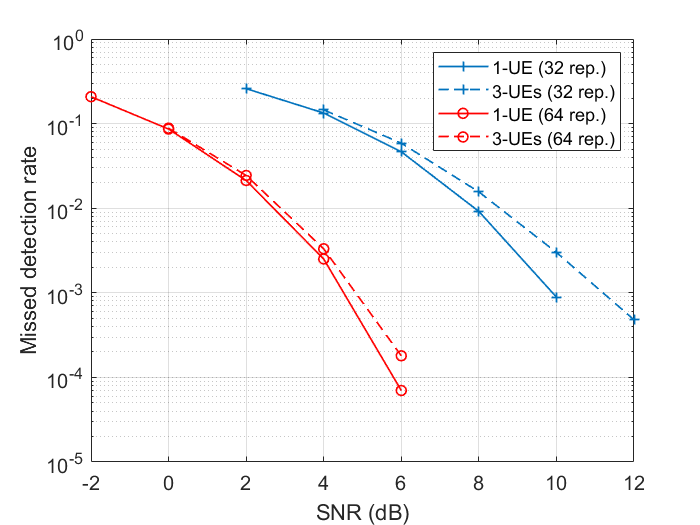}\label{fig:max_RTD_3}}
  \hfill
  \subfloat[NTN beam with MD-RTD of 533.3~$\mu$s]{\includegraphics[width=0.45\textwidth]{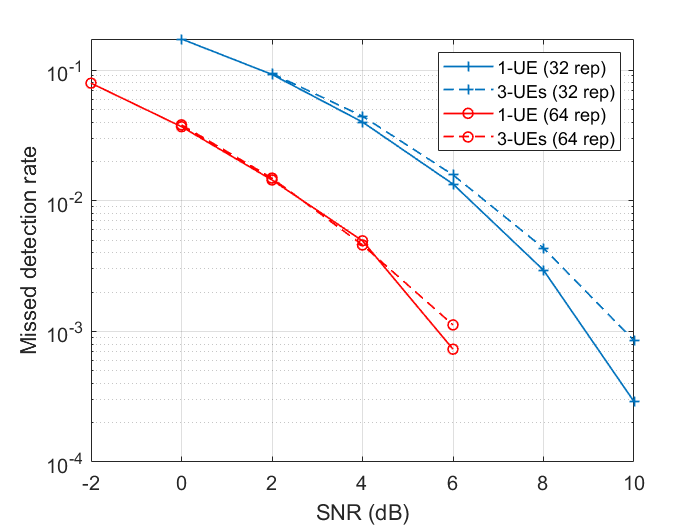}\label{fig:max_RTD_4}}
  \caption{NPRACH preamble detection performance with different MD-RTDs in our NTN scenario.}
  \label{fig:NPRACH_results}
\end{figure*}

\subsection{Obtained results}
\begin{figure}[!tbp]
  \centering
  \subfloat[CDF of absolute Doppler rate estimation error with 32 repetitions]{\includegraphics[width=0.45\textwidth]{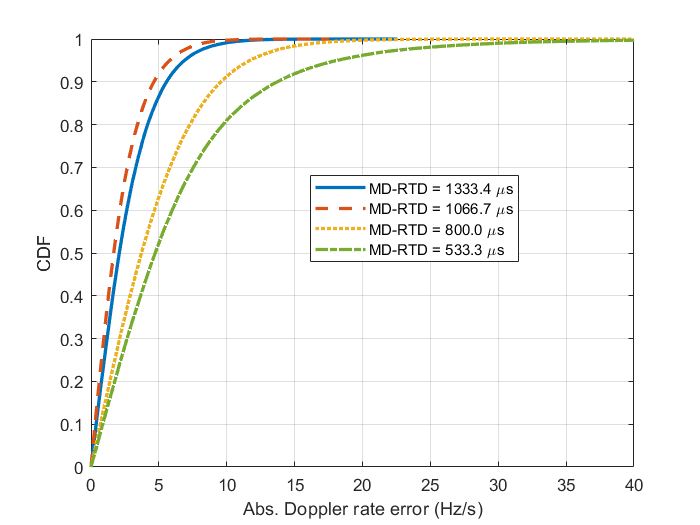}\label{fig:32_rep}}
  \\
  \subfloat[CDF of absolute Doppler rate estimation error with 64 repetitions]{\includegraphics[width=0.45\textwidth]{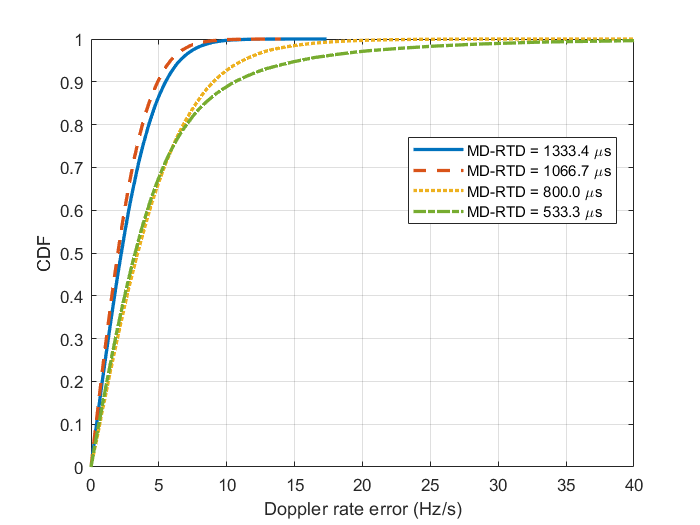}\label{fig:64_rep}}
  \caption{CDF of absolute Doppler rate estimation error at the missed detection target of 10$^{-2}$.}
  \label{fig:CDF_Doppler_error}
\end{figure}

\begin{figure}[!tbp]
  \centering
  \subfloat[CDF of absolute ToA estimation error with 32 repetitions]{\includegraphics[width=0.45\textwidth]{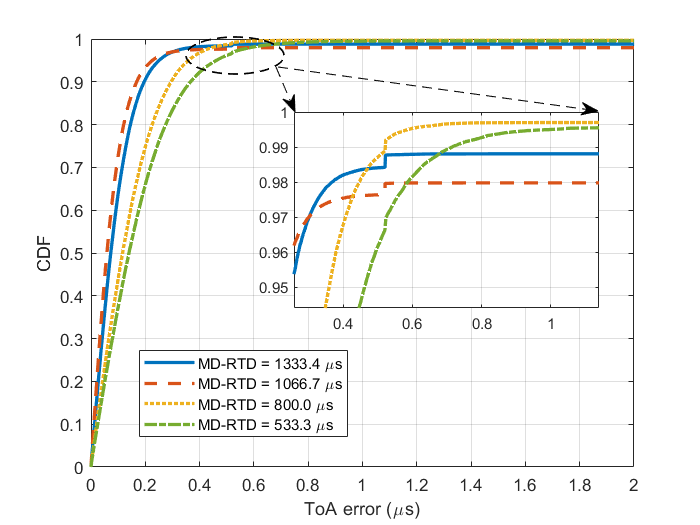}\label{fig:ToA_error_32_rep}}
  \\
  \subfloat[CDF of absolute ToA estimation error with 64 repetitions]{\includegraphics[width=0.45\textwidth]{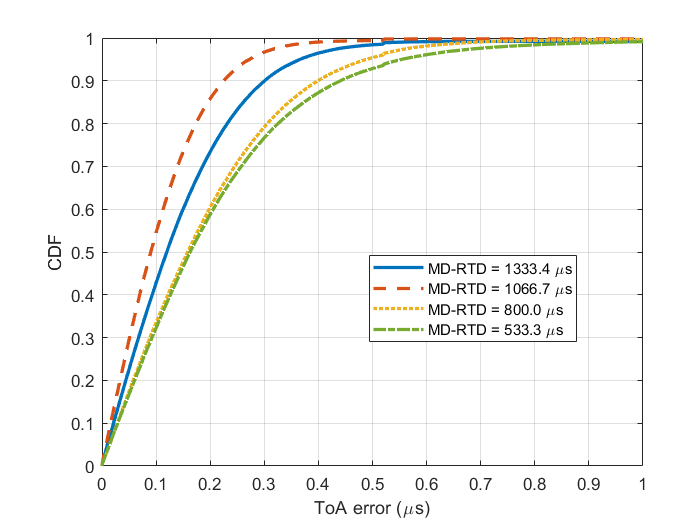}\label{fig:ToA_error_64_rep}}
  \caption{CDF of absolute ToA estimation error at the missed detection target of 10$^{-2}$.}
  \label{fig:CDF_ToA_error}
\end{figure}

The overall obtained results are first analyzed and discussed. The performance of ToA and Doppler rate estimators are provided in terms of cumulative distribution function (CDF) of the absolute ToA estimation error and Doppler rate estimation error.
All results are obtained with a false alarm probability being less than 0.1\%.

\subsubsection{Overall performance}
The obtained results are depicted in Fig.~\ref{fig:NPRACH_results} and presented in terms of missed detection rate versus SNR.
We start first by presenting and analyzing the case of single UE.
For an MD-RTD of 1333.4~$\mu$s (e.g. scenario for beam at minimum elevation angle of 31$^\circ$ for Set-2 configuration with 246-km diameters) the missed detection target of 10$^{-2}$ is achieved at SNR of 15.5~dB and 9.3~dB for 32 and 64 repetitions, respectively (see Fig.~\ref{fig:max_RTD_1}).
In case of an MD -RTD of 1067~$\mu$s (e.g. scenario for beam at minimum elevation angle of 30$^\circ$ for Set-1 configuration with 144-km diameters) the target is reached at SNR of 14.8~dB and 8.1~dB for 32 and 64 repetitions, respectively (see Fig.~\ref{fig:max_RTD_2}).
For lower MD-RTDs, lower SNR values are observed at the missed detection target (see Fig.~\ref{fig:max_RTD_3} and Fig.~\ref{fig:max_RTD_4}). For an MD-RTD of 800~$\mu$s the target is reached at 7.9~dB and 2.8~dB for 32 and 64 repetitions, respectively, whereas for MD-RTD of 533.4~$\mu$s these values are 6.4~dB and 2.6~dB for 32 and 64 repetition, respectively.
When analyzing the results, we notice that the performance is enhanced when we reduce the MD-RTD. This is an expected behavior, since the number of useful symbols ($\notin$ CP) is inversely proportional to the CP length. This leads to SNR enhancement (i.e. performance enhancement) when combining within the same SG in \eqref{eq:combine within Ymi}. Furthermore, this is also related to the fact that at lower elevation angles, where higher D-RTDS are expected, better Doppler rate accuracy is needed compared to higher elevation angles, as explained in Section~\ref{subsec:ToA selection}.

Regarding the case of three active UEs, naturally, we observe some performance degradation compared to the single-UE performance. This degradation is not the same for all cases. For MD-RTDs above 1000~$\mu$s, we observe a relatively substantial degradation before the missed detection target in case of 32 repetitions, and about 0.5-dB degradation at the missed detection target in case of 64 repetitions. On the other hand, for MD-RTDs below 1000~$\mu$s, we notice about 0.5-dB degradation in case of 32 repetitions and negligible degradation in case of 64 repetitions.
The low impact of ICI at lower SNRs is understandable. In fact, at low SNR regime, the noise is the dominant impairment. But when moving to moderate and high SNR regimes, the ICI starts to be the dominant impairment leading to performance degradation.
The substantial degradation in case of 32 repetitions at higher MD-RTDs ($>$ 1000~$\mu$s) can be explained by the lack of signal length (less repetitions) to estimate the Doppler rate with enough accuracy in presence of ICI. On the other hand, it seems sufficient to have 64 repetitions to reach the required Doppler rate accuracy, leading to better overall performance.

\subsubsection{Estimators' performance}
Fig.~\ref{fig:CDF_Doppler_error} and Fig.~\ref{fig:CDF_ToA_error} respectively show the performance of Doppler rate and ToA estimators around the missed detection target of 10$^{-2}$ for both 32 and 64 repetitions when three UEs are active.

First, when observing the Doppler estimation accuracy (see Fig.~\ref{fig:CDF_Doppler_error}), one can notice that the estimation accuracy required to achieve the missed detection target is high in case of large D-RTDs (i.e. lower elevation angles), compared to small D-RTDs (i.e. higher elevation angles) for all repetitions.
For 32 repetitions, the Doppler rate errors (at 99\% confidence level) are 9.8 Hz/s, 8.3 Hz/s, 17 Hz/s, 30.5 Hz/s for MD-RTDs of 1333.4~$\mu$s, 1066.7~$\mu$s, 800~$\mu$s and 533.3~$\mu$s, respectively.
In case of 64 repetitions, those values are 8.7 Hz/s, 7.8 Hz/s, 17 Hz/s, and 31 Hz/s for MD-RTDs of 1333.4~$\mu$s, 1066.7~$\mu$s, 800~$\mu$s and 533.3~$\mu$s, respectively.
As explained in Section~\ref{subsec:ToA selection}, this reflects the fact that for large D-RTD (i.e. lower elevation angles) one expects wider beams, smaller Doppler rates,  and a high number of ToA candidates to be discriminated, thus requiring a very good accuracy compared to higher elevation angles.
In addition, one can notice that the accuracy needed for MD-RTD of 1333.4~$\mu$s is less than the one for MD-RTD of 1066.7~$\mu$s. This is because the latter MD-RTD represents Set-1 configuration, which has a narrower beam compared to the case of MD-RTD of 1333.4~$\mu$s in Set-2 configuration.
Note that for 32 repetitions and in case of MD-RTDs of 1333.4~$\mu$s and 1066.7~$\mu$s, the required accuracy (expected to be $\approx$ 9 Hz/s and $\approx$ 8 Hz/s, respectively) to attain the target missed detection is not reached. This is why in those cases (i.e. 32 repetitions with MD-RTD of 1333.4~$\mu$s and 1066.7~$\mu$s) the missed detection target is not achieved.

Regarding the ToA estimator, as shown in Fig.~\ref{fig:CDF_ToA_error}, it is indeed very accurate at the target missed detection rate. For 32 and 64 repetitions, the accuracy of the estimator is less than 1~$\mu$s except for cases where MD-RTD is equal to 1333.4~$\mu$s and 1066.7~$\mu$s with 32 repetitions.
Those exceptions stem from the aforementioned lack of Doppler rate accuracy, which leads to a poor discrimination among the ToA candidates.

\subsubsection{Concluding remarks on the results}
The overall performance demonstrates the ability of the proposed method to address the NTN scenario described in Section~\ref{sec:System Model}. Explicitly, the method can address UEs at minimum elevation angle of 31$^\circ$ for Set-2 configuration and UEs at 30$^\circ$ elevation angles and even lower for Set-1 configuration.
For both configurations, 64 preamble repetitions are needed for UEs at lower elevation angles, whereas for higher ones, the UEs can be served with lower preamble repetitions.
It is worth noting here that that the obtained results may constitute a new reference for future investigations of NPRACH detection performance in the NTN context, where the available literature is rather scarce.

{\subsection{Discussion}
The proposed method in this paper is designed to work with UEs without GNSS but could also work with UEs integrating GNSS capability. The method can be useful in terms of ToA estimation accuracy by eliminating the CFO as well as the Doppler rate effects when estimating the ToA. Furthermore, independently of the UE type (with or without GNSS), the estimation of Doppler rate can be reported to upper layers for the scheduling algorithm, which could use it to preserve the UEs' orthogonality for data transmission \cite{Kisseleff9237970}.}

{On the other hand, the actual performance of the method depends on the satellite altitude. Satellites at higher altitudes have larger spotbeams and lower speed,  leading to larger D-RTD and lower Doppler rates, assuming the same payload configurations. First, the method (i.e. without GNSS) cannot work if D-RTD exceeds the MD-RTD that the system can handle (i.e. $T_{\rm SEQ}$). This issue can be tackled by considering higher elevation angles (e.g. for a LEO at 2000 km, the minimum elevation angle becomes around 42$^{\circ}$ for Set-1 configuration) for the system operations.
Now, when the Doppler rate is very small, the method may not be able to discriminate between the possible D-RTDs,  since  it  is  limited  by  the  precision  of  the  Doppler  rate  estimator. For  example,  with a LEO at 2000 km and minimum elevation angle of 42 degree with Set-1 configuration, the Doppler rate  variation  between the two edges of the spotbeam is around 12 Hz/sec (assuming  a  maximum Doppler rate  of  125  Hz/sec).  Accordingly,  in  order  to  discriminate  5  D-RTDs, we need a precision better than 12/5/2 = 1.2 Hz/sec, which is difficult to achieve for 99\% of cases. In this configuration, adopting higher carrier frequency may solve this issue.}


\section{Conclusion}\label{Conclusion}
In this work, a new receiver method was designed for the integration of NB-IoT random access in non-terrestrial networks. A detailed scenario definition along with link budget description as well as some design aspects related to the integration of NB-IoT random access via LEO satellites were provided. 
The proposed method is designed to minimize the impact of frequency offset and Doppler rate while extending the coverage beyond the typical limit of the NB-IoT system in terrestrial networks.
Performance evaluation showed how the proposed method can address the different configurations defined by 3GPP standardization body for non-terrestrial networks. Since numerous companies are interested into the integration of NB-IoT via satellite, the proposed method can constitute a practical and seamless solution for the random access channel.

\section*{Appendix~A\\ \quad\ Mathematical symbols used in Preamble Detection and ToA Estimation in both TN and NTN}\label{Appendix: Mathematical symbols}
	\begin{supertabular}{lp{0.375 \textwidth}}
		
				\textbf{\textit{Symbol}} & \textbf{\textit{Definition}} \\
		         $s_{m,i}[n]$   & Transmit signal at the $n$-th sample of the $i$-th symbol in $m$-th symbol group \\
		        $S_{m,i}[k]$  & $i$-th symbol on the $k$-th subcarrier during the $m$-th SG\\
		       $N_{g}$         &  Size of one symbol group in sample\\
		        $N_{\rm CP}$ & CP size in sample\\
		      $N$            & Symbol size in sample\\
		      $y_{m,i}[n]$ & Received time domain signal of the $n$-th sample of the $i$-th symbol in the $m$-th symbol group\\
		      $Y_{m,i}[l]$  & Received frequency domain signal of the $l$-th subcarrier of the $i$-th symbol in the $m$-th symbol group\\
		      $W_{m,i}$ & Additive noise signal in the $i$-th symbol in the $m$-th symbol group\\
		     $Y_{m}$      & Sum of frequency domain symbols of the $m$-th symbol group \\
		     $L_{\rm CP}$ & Number of symbols employed to extend the CP within one symbol group\\
		     $L^\prime$ & Number of remaining symbols after CP removal \\
			$f_{\rm off}$	 & Carrier frequency offset normalized by the sampling frequency \\
			 $D$   & RTD normalized by the symbol duration\\
			 $\alpha$ & Doppler rate normalized by squared sampling frequency\\
			 $\alpha_{\max}$ & Maximum Doppler rate that can be theoretically estimated\\
			 $\alpha_{\max}^{\rm sc}$ & Maximum Doppler rate for the considered scenario\\
			 $\alpha_{\min}^{\rm sc}$ & Minimum Doppler rate for the considered scenario\\
			 
			 $h_{m,i}$  & Channel gain of the $i$-th symbol in the $m$-th symbol group.\\
				$h_{m}$  & Channel gain of the  the $m$-th symbol group\\
				$n_{\rm SC}^{\rm RA}(m)$      & Subcarrier occupied by the $m$-th symbol group\\
             $\Delta(m)$   & Hopping step between the $m$-th and $(m+1)$-th symbol groups\\
             $\Delta(m)_{min}$   & Minimum hopping step in the preamble\\
               $Z_{m,1}$ & Differential symbol resulting from $m$-th and $(m+1)$-th symbol group differential processing\\
                $v$ & Array containing differential symbols with corrected frequency hopping \\
                $v_{u}$ & Array containing differential symbols of $u$-th preamble basic unit with corrected frequency hopping\\
               $N_{\rm DFT}$ & Number of DFT points \\
               $N_{\rm RX}$ & Number of receive antennas\\
               $V$ & Post-DFT vector of corrected differential symbols\\
               $V_{u}$ & Post-DFT vector of corrected differential symbols of the $u$-th preamble basic unit\\
              $X[k]$  & NPRACH detection metric vector \\
              $X_{\max}$ & Maximum value of the NPRACH detection metric vector\\
               $\hat{D}$ & Estimated ToA (i.e. RTD)\\
               $t_{m}$ & ToA corrected differential symbol\\
               $T$ & Post-DTFT vector of ToA corrected differential symbols\\
               $N_{t}$ & Number of symbols in $t$ vector\\
               $N_{\rm DTFT}$ & Number of DTFT points\\
               $J$ & Post-DTFT vector of corrected differential symbols summed across receive antennas\\
               $\hat{\alpha}$ & Estimate of Doppler rate\\
              $ \delta_{\alpha}$ & Resolution of Doppler rate estimation\\
		\end{supertabular}
  		
\color{black}
\bibliographystyle{IEEEtran}
\bibliography{main.bib}

\begin{thebibliography}{10}
\providecommand{\url}[1]{#1}
\csname url@samestyle\endcsname
\providecommand{\newblock}{\relax}
\providecommand{\bibinfo}[2]{#2}
\providecommand{\BIBentrySTDinterwordspacing}{\spaceskip=0pt\relax}
\providecommand{\BIBentryALTinterwordstretchfactor}{4}
\providecommand{\BIBentryALTinterwordspacing}{\spaceskip=\fontdimen2\font plus
\BIBentryALTinterwordstretchfactor\fontdimen3\font minus
  \fontdimen4\font\relax}
\providecommand{\BIBforeignlanguage}[2]{{%
\expandafter\ifx\csname l@#1\endcsname\relax
\typeout{** WARNING: IEEEtran.bst: No hyphenation pattern has been}%
\typeout{** loaded for the language `#1'. Using the pattern for}%
\typeout{** the default language instead.}%
\else
\language=\csname l@#1\endcsname
\fi
#2}}
\providecommand{\BIBdecl}{\relax}
\BIBdecl

\bibitem{NGMN}
{NGMN Alliance}, ``{Non-Terrestrial Networks Position Paper},'' \emph{project:
  Extreme Long-Range Communications for Deep Rural Coverage}, 2019-11.

\bibitem{3GPP38913v14}
{$3^{rd}$ Generation Partnership Project}, ``{5G; Study on Scenarios and
  Requirements for Next Generation Access Technologies},'' \emph{3GPP TR 38.913
  version 14.2.0 Release 14}, 2017-05.

\bibitem{Olt9210567}
O.~{Kodheli}, E.~{Lagunas}, N.~{Maturo}, S.~K. {Sharma}, B.~{Shankar}, J.~F.~M.
  {Montoya}, J.~C.~M. {Duncan}, D.~{Spano}, S.~{Chatzinotas}, S.~{Kisseleff},
  J.~{Querol}, L.~{Lei}, T.~X. {Vu}, and G.~{Goussetis}, ``Satellite
  communications in the new space era: A survey and future challenges,''
  \emph{IEEE Communications Surveys Tutorials}, pp. 1--1, 2020.

\bibitem{3GPP38811v0}
{$3^{rd}$ Generation Partnership Project}, ``{3rd Generation Partnership
  Project; Technical Specification Group Radio Access Network; Study on New
  Radio (NR) to support Non Terrestrial Networks (Release 15)},'' \emph{3GPP TR
  38.811V0.1.0}, 2017-06.

\bibitem{3GPP38821v16}
------, ``{3rd Generation Partnership Project; Technical Specification Group
  Radio Access Network; Solutions for NR to support non-terrestrial networks
  (NTN) (Release 16)},'' \emph{3GPP TR 38.821 V16.0.0}, 2019-12.

\bibitem{3GPP38811v15}
------, ``{3rd Generation Partnership Project; Technical Specification Group
  Radio Access Network; Study on New Radio (NR) to support Non Terrestrial
  Networks (Release 15)},'' \emph{3GPP TR 38.811V15.1.0}, 2019-06.

\bibitem{RP193144thales}
{Thales}, ``{Solutions for NR to support non-terrestrial networks (NTN)},''
  \emph{{RP-193144 3GPP TSG RAN meeting \#86}}, 2019-12.

\bibitem{RP193235MediaTek}
{MediaTek Inc.}, ``{New Study WID on NB-IoT/eTMC support for NTN },''
  \emph{RP-193235 3GPP TSG RAN meeting \#86}, 2019-12.

\bibitem{lin2016random}
X.~Lin, A.~Adhikary, and Y.-P. Wang, ``{Random access preamble design and
  detection for 3GPP narrowband IoT systems},'' \emph{IEEE Wireless
  Communications Letters}, vol.~5, no.~6, pp. 640--643, 2016.

\bibitem{3GPP36211}
{$3^{rd}$ Generation Partnership Project}, ``{LTE; Evolved Universal
  Terrestrial Radio Access (E-UTRA); Physical channels and modulation},''
  \emph{3GPP TS 36.211 version 13.7.1 Release 13}, 2017-10.

\bibitem{3GPP36331}
------, ``{LTE; Evolved Universal Terrestrial Radio Access (E-UTRA); Radio
  Resource Control (RRC); Protocol specification},'' \emph{3GPP TS 36.331
  version 13.8.1 Release 13}, 2018-01.

\bibitem{charbit2020space}
G.~Charbit, D.~Lin, K.~Medles, L.~Li, and I.-K. Fu, ``{Space-Terrestrial Radio
  Network Integration for IoT},'' in \emph{2020 2nd 6G Wireless Summit (6G
  SUMMIT)}.\hskip 1em plus 0.5em minus 0.4em\relax IEEE, 2020, pp. 1--5.

\bibitem{kodheli2020random}
O.~Kodheli, N.~Maturo, S.~Chatzinotas, S.~Andrenacci, and F.~Zimmer, ``{On the
  Random Access Procedure of NB-IoT Non-Terrestrial Networks},'' 2020.

\bibitem{ESA9149179}
A.~{Guidotti}, A.~{Vanelli-Coralli}, A.~{Mengali}, and S.~{Cioni},
  ``Non-terrestrial networks: Link budget analysis,'' in \emph{ICC 2020 - 2020
  IEEE International Conference on Communications (ICC)}, 2020, pp. 1--7.

\bibitem{R11907481ESA}
{ESA}, ``{Beam size computation and alternative satellite specifications},''
  \emph{R1-1907481 3GPP TSG WG1 Meeting \#97}, 2019-05.

\bibitem{R1-1913224ESA}
------, ``{NTN system level and link budget calibration: DL and UL results},''
  \emph{R1-1913224 3GPP TSG RAN WG1 Meeting \#99}, 2019-11.

\bibitem{ericsson2015}
{Ericsson}, ``{Narrowband LTE – random access design},'' \emph{R1-156011,
  3GPP TSG-RAN1 \#82bis}, October 2015.

\bibitem{ericsson2015x2}
------, ``{NB-IoT – random access design},'' \emph{R1-157424, 3GPP
  TSG-RAN1\#83}, November 2015.

\bibitem{ericsson2016}
------, ``{NB-IoT – design considerations for single tone frequency hopped
  NB-PRACH},'' \emph{R1-160093, 3GPP TSG-RAN1 AH-NB-IoT}, January 2016.

\bibitem{ericsson2016x2}
------, ``{NB-IoT – single tone frequency hopping NB-PRACH design},''
  \emph{R1-160275, 3GPP TSG-RAN1 \#84}, February 2016.

\bibitem{3GPP36300}
{$3^{rd}$ Generation Partnership Project}, ``{LTE; Evolved Universal
  Terrestrial Radio Access (E-UTRA) and Evolved Universal Terrestrial Radio
  Access Network (E-UTRAN); Overall description},'' \emph{3GPP TS 36.300
  version 13.2.0 Release 13}, 2016-01.

\bibitem{3GPP36211v15}
------, ``{LTE; Evolved Universal Terrestrial Radio Access (E-UTRA); Physical
  channels and modulation},'' \emph{3GPP TS 36.211 version 15.2.0 Release 15},
  2018-10.

\bibitem{Houcine2020}
H.~{Chougrani}, S.~{Kisseleff}, and S.~{Chatzinotas}, ``{Efficient Preamble
  Detection and Time-of-Arrival Estimation for Single-Tone Frequency Hopping
  Random Access in NB-IoT},'' \emph{IEEE Internet of Things Journal}, pp. 1--1,
  2020, 10.1109/JIOT.2020.3039004.

\bibitem{R1-1908049Huawei}
{Huawei, HiSilicon, CAICT}, ``{Discussion on Doppler compensation, timing
  advance and RACH for NTN},'' \emph{R1-1908049 3GPP TSG RAN WG1 Meeting \#98},
  2019-08.

\bibitem{morelli2002doppler}
M.~Morelli, ``Doppler-rate estimation for burst digital transmission,''
  \emph{IEEE Transactions on Communications}, vol.~50, no.~5, pp. 707--710,
  2002.

\bibitem{mengali2013synchronization}
U.~Mengali, \emph{Synchronization techniques for digital receivers}.\hskip 1em
  plus 0.5em minus 0.4em\relax Springer Science \& Business Media, 2013.

\bibitem{3GPP36104}
{$3^{rd}$ Generation Partnership Project}, ``{LTE; Evolved Universal
  Terrestrial Radio Access (E-UTRA); Base Station (BS) radio transmission and
  reception},'' \emph{3GPP TS 36.104 version 14.3.0 Release 14}, 2017-04.

\bibitem{Ericsson160277}
{Ericsson}, ``{NB-IoT – Near-Far Performance of NB-PRACH},'' \emph{{R1-160277
  3GPP TSG-RAN1 \#84}}, 2016-02.

\bibitem{Kisseleff9237970}
S.~{Kisseleff}, E.~{Lagunas}, T.~S. {Abdu}, S.~{Chatzinotas}, and
  B.~{Ottersten}, ``Radio resource management techniques for multibeam
  satellite systems,'' \emph{IEEE Communications Letters}, pp. 1--1, 2020.

\end{thebibliography}
\end{document}